\documentclass[12pt]{article}

\usepackage[centertags]{amsmath}
\usepackage{amsfonts}
\usepackage{amssymb}
\usepackage{amsthm}
\usepackage{newlfont}
\usepackage{epsfig}
\usepackage{amscd}

\newcommand{\NN}{{\mathbb N}}

\newcommand{\RR}{{\mathbb R}}

\newcommand{\CC}{{\mathbb C}}
\newcommand{\beq}{\begin{equation}}
\newcommand{\eeq}{\end{equation}}
\newcommand{\ba}{\begin{array}}
\newcommand{\ea}{\end{array}}
\newcommand{\bea}{\begin{eqnarray}}
\newcommand{\eea}{\end{eqnarray}}
\newcommand{\eps}{\epsilon}

\begin{document}

\begin{center}
{\large \sc \bf Multiscale expansions of difference equations \\ in the small lattice spacing regime, and \\  
a vicinity and integrability test. I} 

\vskip 20pt

{\large  Paolo Maria Santini$^{1,\S}$}

\vskip 20pt

{$^1$ Dipartimento di Fisica, Universit\`a di Roma "La Sapienza", and \\
Istituto Nazionale di Fisica Nucleare, Sezione di Roma 1 \\
Piazz.le Aldo Moro 2, I-00185 Roma, Italy}

\bigskip

$^{\S}$e-mail:  {\tt paolo.santini@roma1.infn.it}

\bigskip

{\today}

\end{center}

\begin{abstract}
We propose an algorithmic procedure i) to study the ``distance'' between an integrable PDE and 
any discretization of it, in the small lattice spacing $\eps$ regime, and, at the same time, ii) to test the 
(asymptotic) integrability properties of such discretization. This method should provide, in particular,  
useful and concrete informations on how good is any numerical scheme used to integrate a given integrable PDE. 
The procedure, illustrated on a fairly general 10-parameter family of discretizations of the nonlinear 
Schr\"odinger equation, consists of the following three steps: i) the construction of 
the continuous multiscale expansion of a generic solution of the discrete system at all orders in $\eps$, following \cite{DMS}; 
ii) the application, to such expansion, of the Degasperis - Procesi (DP) integrability test \cite{DP,Degasperis}, to test 
the asymptotic integrability properties of the discrete system and its ``distance'' from its continuous limit; 
iii) the use of the main output of the DP test to construct infinitely many approximate symmetries and constants of motion 
of the discrete system, through novel and simple formulas. 
 
\end{abstract}
\section{Introduction}

Given a partial differential equation (PDE) and a partial difference equation (P$\Delta$E) discretizing it, 
it is interesting to know, when the lattice spacing $\eps$ is small, 
``how close'' the two models are. In particular, if the PDE is integrable, it is important to have a way to 
establish if such a discretization preserves integrability or, at least, how ``close'' is to an integrable system,  
detecting the order, in $\eps$, at which the discretization departs from integrability and, correspondingly, the time scale 
at which one should expect numerical evidence of nonintegrability and/or chaos. In addition, given a PDE and two 
P$\Delta$Es discretizing it, it is also interesting to know, when the lattice spacing $\eps$ is small, 
``how close'' the two P$\Delta$Es are.    

In this paper we propose to answear these basic questions in the following way. Concentrating on an integrable PDE  
and on a P$\Delta$E discretizing it, \\
1) we construct and study in detail the multiscale expansion at all orders of a {\it generic solution} of the P$\Delta$E 
under scrutiny, generated in the small $\eps$ regime,  following the procedure 
developed in \cite{DMS}. At $O(1)$, the leading term $u$ of such asymptotic expansion satisfies the integrable PDE;  
to keep the expansion asymptotic, we eliminate the secularities due to the linear part of the P$\Delta$E, arising at each order,  
introducing infinitely many slow (time) variables and establishing that the evolution of $u$ with respect to such slow times is 
described by the infinite hierarchy of commuting flows of the integrable PDE, as in \cite{DMS}.         \\
2) We make use of the asymptotic 
integrability test developed by Degasperis-Procesi (DP) in \cite{DP,Degasperis} on such a multiscale expansion to test, 
at all orders, the 
``asymptotic'' integrability properties of the P$\Delta$E; in particular, detecting the order in $\eps$ 
(and, correspondingly, the time scale) at which the discretization departs from integrability. At this time scale, f.i., 
numerical simulations are expected to give some evidence of non-integrable and/or chaotic behaviour. \\
3) We finally show how to make use of the main output of the DP test to construct infinitely many 
``approximate'' symmetries, at a required order in $\eps$, of the P$\Delta$E under scrutiny, using novel and simple formulas.

Recent studies on the performances, as numerical schemes for their continuous limits, of P$\Delta$Es possessing the same 
(continuous) Lie point symmetries as their continuous limits can be found in \cite{W,VW}. Studies on the performances, as 
numerical schemes for their continuous limits, of integrable discretizations of 
integrable PDEs can be found, f. i., in \cite{TA} and \cite{HA}; in this case, the integrable discretization possesses 
infinitely many exact generalized symmetries and constants of motion in involution at any  
order in $\eps$, reducing to the generalized symmetries and constants of motion of the integrable PDE in the 
continuous limit. The P$\Delta$Es selected by our approach 
possess instead infinitely many approximate generalized symmetries and constants of motion in involution at the 
required order in $\eps$ (see \S 3.1), reducing to the generalized symmetries and constants of motion of the integrable PDE in the 
continuous limit.    
 
The procedure we propose should allow one to have a control on the ``distance'' between the P$\Delta$E and its continuous limit, 
as well as on the distance between two different discretizations of the same PDE. Indeed,   
suppose we construct an asymptotic expansion of the form $\psi=u+O(\eps^{\alpha}),~\alpha>0$, where $\psi$ is a generic 
solution of the P$\Delta$E and $u$ is the corresponding solution of its continuous limit; if, at $O(\eps^{\beta})$, $\beta>0$,  
the P$\Delta$E passes the DP test, we infer that $||\psi-u||=O(\eps^{\alpha})$ at time scales of 
$O(\eps^{-\beta})$, where $||\cdot ||$ is the uniform norm wrt $x$ and $t$ (the norm used to test the asymptotic character of the 
generated multiscale expansion). In this way, since we control the distance between 
``generic solutions'' of the P$\Delta$E and of its continuous limit, we also control the distance 
between the P$\Delta$E and its continuous limit. In addition, if the multiscale expansions of 
two different discretizations of the same PDE pass the DP test at $O(\eps^{\beta})$, we infer, from the triangular inequality, that 
$||\psi-\phi ||<||\psi-u||+||\phi-u||=O(\eps^{\alpha})$ at time scales of $O(\eps^{-\beta})$, where $\psi,~\phi$ 
are solutions of the two different discretizations of the PDE corresponding to the same generic initial-boundary data; 
therefore we have a control also on the distance between the two different discretizations of the same PDE. 

Some historical remarks are important, at this point, on the theory of multiscale expansions in connection with 
integrable systems, to put the results of this paper into a proper perspective. 
Multiscale expansions of a given PDE are very useful tools for investigating the properties 
of such a PDE and for identifying important model (universal) equations of physical phenomena. 
For instance, if the original nonlinear PDE has a dispersive linear part, a small amplitude monochromatic wave 
evolving according to it develops a slow space-time amplitude modulation described 
by the celebrated nonlinear Schr\"odinger (NLS) equation \cite{Kelley,Zakharov,BN,HT,HO} (see also \cite{Taniuti,KT,CE1})    
\beq\label{nls1}
iu_t+u_{xx}+2c|u|^2u=0,~~u=u(x,t)\in\CC,
\eeq
integrable if $c$ is a real constant \cite{ZS}. Considering, instead, three monochromatic waves  
and imposing a suitable resonance condition on their wave numbers and dispersion relations, one generates another 
integrable universal model, the 3-wave resonant system \cite{ZM}. In the above two examples, the expansion is 
constructed around ``approximate'' particular solutions of the original PDE (the monochromatic waves). It is also 
possible to 
expand around ``exact'' particular solutions of the original PDE; for instance, as shown in \cite{ZK}, expanding  
around the exact solution $u_0=exp(2ict)$ of (\ref{nls1}), the first nontrivial term of the asymptotic expansion evolves 
according to another important model equation: the Korteweg-de Vries (KdV) equation \cite{KdV}, sharing with NLS 
the property of integrability \cite{GGKM}.     
Since multiscale expansions preserve integrabilty \cite{ZK}, i) if the original PDE is a ``C-integrable'' system  
(i.e., it is linearized by a ``change of variables'' \cite{CE1,Calogero}, like the Burgers equations \cite{Hopf-Cole}), 
the model equation generated by it is linear \cite{CE1,Calogero}; ii) if the original PDE is an ``S-integrable'' system, or soliton 
equation (like the NLS equation), integrated in a more complicated way via a Riemann-Hilbert or $\bar\partial$-problem 
\cite{ZMNP,CD,AC,Konop}, the model equation generated by it is also ``S-integrable''; viceversa, iii) if the model equation generated 
by the expansion is not integrable, then the original equation 
is not integrable too. This criterion has been used in \cite{CE1,CE2,CDJ1,CDJ2} as a simple test of integrability.   
In addition, the universal character of the identified model equations  
(NLS, KdV or others) is also the reason why model equations possess very distinguished mathematical properties and, often, 
they are integrable \cite{CE1,CE2,Calogero}.  

Multiscale expansions can also be carried, in principle, to all orders and, 
as a consequence of eliminating the secular terms at each order, a sequence of  
slow time variables $t_n=\eps^n t$ must be introduced and the dependence of the leading term of the 
expansion on such slow times is described by the hierarchy of commuting flows of the integrable model equation   
\cite{DMS}. This multiscale expansion at all orders has been used in \cite{DP, Degasperis} to build an efficient 
asymptotic integrability test for the original PDE (see \S 3 for more details on such test). An alternative asymptotic 
integrability test, based 
on the existence of approximate symmetries for the original PDE, can be found in \cite{KM}. 
The ideas and procedures developed in \cite{DMS,DP, Degasperis} have been recently used to build an 
integrability test also for P$\Delta$Es \cite{Levi,Tesi,Scimi}; in this approach, one expands, as for the PDE case, around 
approximate or exact particular solutions of the P$\Delta$E under investigation, obtaining a continuous multiscale 
expansion at all orders, following \cite{DMS}, and applying on it the DP test.  
The main difference between the procedure followed in \cite{Levi,Tesi,Scimi} and the results of this paper is the following. 
The standard multiscale approach used in \cite{Levi,Tesi,Scimi}, obtained expanding 
around approximate or exact particular solutions of the P$\Delta$E under investigation, cannot give informations 
on how close this P$\Delta$E and its continuous limit are, the main goal of the present paper. The common features of 
the procedure in \cite{Levi,Tesi,Scimi} and of that used in this paper are that, in both cases, 
one constructs, from the given P$\Delta$E, continuous multiscale expansions carried to all orders, 
as in \cite{DMS}, and one applies to them the DP integrability test. Therefore both procedures can be used to test the 
integrability and the asymptotic integrability of the original P$\Delta$E.
 
Another integrability test for P$\Delta$Es is the so-called ``symmetry approach'' \cite{Yam}, 
based on the existence of higher order symmetries and originally developed to test the integrability of  
PDEs \cite{SS,MSY}. 

The results of this paper are illustrated on the basic prototype example of the NLS equation (\ref{nls1}), starting from   
the following discretization of it:
\beq\label{dnlsF}
\ba{l}
i\psi_{n,t}+\eps^{-2}\left(\psi_{n+1}+\psi_{n-1}-2\psi_n\right)+F(\psi_{n-1},\psi_n,\psi_{n+1})=0,  \\
~~ \\
F(\psi_{n-1},\psi_n,\psi_{n+1}):=2 a_1|\psi_n|^2\psi_n+a_2 |\psi_n|^2(\psi_{n+1}+\psi_{n-1})+ \\
a_3 \psi^2_n (\bar\psi_{n+1}+\bar\psi_{n-1})+a_4 \psi_n (|\psi_{n+1}|^2+|\psi_{n-1}|^2)+  \\
a_5 \psi_n (\bar\psi_{n+1}\psi_{n-1}+\psi_{n+1}\bar\psi_{n-1})+
a_6\bar\psi_{n}(\psi^2_{n+1}+\psi^2_{n-1})+ \\
2 a_7\bar\psi_{n}\psi_{n+1}\psi_{n-1}+a_8(|\psi_{n+1}|^2\psi_{n+1}+|\psi_{n-1}|^2\psi_{n-1})+ \\
a_9(\psi^2_{n+1}\bar\psi_{n-1}+\psi^2_{n-1}\bar\psi_{n+1})+a_{10}(|\psi_{n+1}|^2\psi_{n-1}+|\psi_{n-1}|^2\psi_{n+1}),
\ea 
\eeq 
where the constant coefficients $a_j,~j=1,\dots,10$ are real, reducing to (\ref{nls1}) in the natural continuous limit 
in which the lattice spacing $\epsilon\to 0$ and $n\epsilon\to x\in\RR$, $\psi_n(t)\to u(x,t)$, with 
\beq\label{c}
c=\sum\limits_{j=1}^{10}a_j.
\eeq 

The $10$-parameter family of equations (\ref{dnlsF}) 
has been recently taken in \cite{Peli} as the starting point of an analysis 
devoted to the identification of discretizations of NLS that possess, at the same time, a solitary wave and a breather 
solution reducing, respectively, to the one soliton and breather solutions of the NLS equation (\ref{nls1}), 
in the continuous limit $\eps\to 0$. We remark that, rescaling the dependent variable, one can always 
introduce one normalization for the $10$ coefficients; for instance, one can choose one of these 
coefficients, say $a_j$, to be $sign(a_j)$ or, better for our purposes, one can normalize the sum (\ref{c}) of the 10 coefficients   
to coincide with the prescribed coefficient $c$ of the NLS equation (\ref{nls1}).   
 
The linear part of the discrete NLS (dNLS) (\ref{dnlsF}) is the standard discretization 
of $(iu_t+u_{xx})$; its nonlinear part is uniquely fixed by the following, physically sound, properties \cite{Peli}. a) 
Equation (\ref{dnlsF}) must possess the gauge symmetry of first kind (i.e., if $\psi_n$ is a solution, 
$\psi_ne^{-i\theta}$ is 
a solution too, where $\theta$ is an arbitrary real parameter), corresponding to the 
infinitesimal gauge symmetry $-i\psi_n$. b) The nonlinearity is cubic; i.e., is the weakest nonlinearity 
compatible with the above gauge symmetry. c) Only first neighbours interactions are considered. d) 
Equation (\ref{dnlsF}) is invariant under the symmetry transformation $\psi_{n\pm 1}\to \psi_{n\mp 1}$ (space isotropy). 

The dNLS (\ref{dnlsF}) contains, in particular:      \\
1) the integrable Ablowitz - Ladik (AL) equation \cite{AL} 
\beq\label{AL}
\ba{l}
i\psi_{n,t}+\eps^{-2}\left(\psi_{n+1}+\psi_{n-1}-2\psi_n\right)+a_2|\psi_n|^2(\psi_{n+1}+\psi_{n-1})=0 ,
\ea 
\eeq  
for $a_j=a_2\delta_{j2},~j=1,\dots,10$;                \\
2) the discretization 
\beq\label{dNLS1}
i\psi_{n,t}+\eps^{-2}\left(\psi_{n+1}+\psi_{n-1}-2\psi_n\right)+2a_1|\psi_n|^2\psi_n=0,  
\eeq
for $a_j=a_1\delta_{j1},~j=1,\dots,10$, relevant in several applications \cite{Davy,SSH,ELS,HeT,ABDKS},  
whose nonintegrability has been recently shown in \cite{Levi,Tesi} using the DP test;       \\ 
3) the discretization corresponding to  
\beq\label{Peli}
a_{10}=a_8,~~a_1=a_4=a_5=a_6=a_7=a_9=0,
\eeq
with $a_2,a_3,a_8$ arbitrary, possessing a solitary wave as well as a breather 
solution reducing, respectively, to the one soliton and breather solutions of the NLS equation  
in the limit $\eps\to 0$ \cite{Peli};                \\
4) the discretization corresponding to 
\beq\label{OJE}
a_8=a_3,~a_2=2 a_3,~ a_4=2 a_6,~~a_5=a_7=a_9=a_{10}=0,  
\eeq
where $a_1,a_3,a_6$ are given in terms of physical quantities, describing coupled optical waveguides 
embedded in a material with Kerr nonlinearities \cite{OJE}; \\
5) the discretization corresponding to 
\beq\label{CKKS}
a_4=a_2,~~a_1=a_3=a_6=a_8=a_2/2,~~a_5=a_7=a_9=a_{10}=0
\eeq 
(a particular case of (\ref{OJE})), appearing in the modelling of the Fermi-Pasta-Ulam problem \cite{CKKS}.

For special values of the coefficients $a_j$'s the dNLS equation (\ref{dnlsF}) is Hamiltonian. For instance, 
equations (\ref{AL}), (\ref{dNLS1}), (\ref{OJE}) and (\ref{CKKS}) are Hamiltonian \cite{OJE}.   

If $0<\eps<<1$, the discrete scheme (\ref{dnlsF}) approximates the NLS equation (\ref{nls1}),(\ref{c}) 
with an error of $O(\eps^2)$.   
To study more precisely how close equations (\ref{dnlsF}) and (\ref{nls1}) are and, in particular, the integrability 
properties of (\ref{dnlsF}), in this paper we follow the procedure indicated in the first part of this introduction, obtaining the 
following results.
\vskip 5pt
\noindent
1) Due to the structure of the vector field in (\ref{dnlsF}), the generated $\eps$-expansion contains only even powers.  
At $O(\eps^2)$, the dNLS (\ref{dnlsF}) passes the DP test iff the 10 coefficients satisfy 
the elegant quadratic constraint
\beq\label{constr1a}
(a_1-3a_3-2a_4-6a_5-5 a_6 +3a_7-5a_8-13a_9-a_{10})(\sum\limits_{j=1}^{10}a_j)=0,
\eeq 
factorized into two linear constraints.   
If the first constraint $\sum\limits_{j=1}^{10}a_j=0$ is satisfied, we are in the C-integrability framework and the dNLS 
(\ref{dnlsF}) approximates the linear Schr\"odinger 
equation with an error of $O(\eps^2)$, for time scales of $O(\eps^{-2})$. If, instead, the second constraint is satisfied:
\beq\label{Sint}
a_1-3a_3-2a_4-6a_5-5 a_6+3a_7-5a_8-13a_9-a_{10}=0,
\eeq
we are in the S-integrability framework and the dNLS (\ref{dnlsF}) approximates the NLS equation (\ref{nls1}),(\ref{c}) with an 
error of $O(\eps^2)$, for time scales of $O(\eps^{-2})$. 

We remark that, among the ten single dNLS equations obtained choosing only one of the ten 
coefficients different from zero in (\ref{dnlsF}), only the AL equation (\ref{AL}) satisfies the constraint (\ref{constr1a}) and 
passes the test at O($\eps^2$).   
\vskip 5pt
\noindent
2) At $O(\eps^4)$ we have the following two scenarios. In the C-integrability framework, the dNLS (\ref{dnlsF}) approximates, 
with an error of $O(\eps^2)$, the linear Schr\"odinger equation for time scales of $O(\eps^{-4})$ iff the following four linear 
constraints 
\beq\label{Cint_3}
\ba{l}
\sum\limits_{j=1}^{10}a_j=0,~~~~~a_1+a_2 + a_6 + a_7=0,~~~~~a_4 -a_5 + 2 a_8 - 2 a_9=0,                  \\
a_2+2(a_3 + 3 a_5 + 3 a_6 - a_7 + a_8 + 7 a_9)=0
\ea
\eeq
are satisfied by the coefficients. Since one of the real $a_j$'s can always be fixed rescaling the dependent variable $\psi$, 
equations (\ref{Cint_3}) characterize a $5$-parameter family of discrete NLS equations (\ref{dnlsF}) passing the test at 
such a high order. 
 
In the S-integrability framework, the dNLS (\ref{dnlsF}) approximates, with an error of 
$O(\eps^2)$, the NLS equation (\ref{nls1}),(\ref{c}) for time scales of $O(\eps^{-4})$,  
iff the coefficients satisfy, together with the linear constraint (\ref{Sint}),  
the five quadratic constraints (\ref{S4}),(\ref{defQ1})-(\ref{defQ5}). Since these five constraints do not contain the 
term $(a_2)^2$, they are trivially satisfied by the integrable AL equation (\ref{AL}), as it has to be. 
In general we do not expect a parametrization of such 
constraints in terms of elementary functions; however we have been able to construct the following two explicit  
examples of dNLS equations 
\beq\label{dNLS_S_O4_A}
\ba{l}
i\psi_{n,t}+\eps^{-2}\left(\psi_{n+1}+\psi_{n-1}-2\psi_n\right)+a_6\Big(
-8 |\psi_n|^2\psi_n+\frac{4}{3} |\psi_n|^2(\psi_{n+1}+\psi_{n-1})+   \\
4 \psi^2_n (\bar\psi_{n+1}+\bar\psi_{n-1})-4 \psi_n (\bar\psi_{n+1}\psi_{n-1}+\psi_{n+1}\bar\psi_{n-1})+
\bar\psi_{n}(\psi^2_{n+1}+\psi^2_{n-1})+ \\
-2 \bar\psi_{n}\psi_{n+1}\psi_{n-1}\Big)=0,
\ea
\eeq 
\beq\label{dNLS_S_O4_B}
\ba{l}
i\psi_{n,t}+\eps^{-2}\left(\psi_{n+1}+\psi_{n-1}-2\psi_n\right)+a_9\Big(
-48 |\psi_n|^2\psi_n-8 \psi_n (|\psi_{n+1}|^2+\\
|\psi_{n-1}|^2)-8 \psi_n (\bar\psi_{n+1}\psi_{n-1}+\psi_{n+1}\bar\psi_{n-1})+
10 \bar\psi_{n}(\psi^2_{n+1}+\psi^2_{n-1})+ \\
-4 \bar\psi_{n}\psi_{n+1}\psi_{n-1}-7 (|\psi_{n+1}|^2\psi_{n+1}+|\psi_{n-1}|^2\psi_{n-1})+ \\
(\psi^2_{n+1}\bar\psi_{n-1}+\psi^2_{n-1}\bar\psi_{n+1})+6(|\psi_{n+1}|^2\psi_{n-1}+|\psi_{n-1}|^2\psi_{n+1})\Big)=0,
\ea
\eeq
satisfying such complicated quadratic constraints, corresponding to particular cases in which the associated   
five quadrics degenerate into hyperplanes.

These two distinguished models, passing the test at such a high order through the above degeneration mechanism, are obviously 
good candidates to be S-integrable discretizations of NLS. A detailed study of their performances as numerical schemes 
for NLS, and of their possible integrability structure (Lax pair, etc ..) is postponed to a subsequent paper.  
    
To obtain the above results, it is essential to use  
well-known integrability properties of equation (\ref{nls1}) ( shared by all integrable systems;  
see, f.i. \cite{Magri}-\cite{SF}) that we summarize here, for completeness. 

The NLS equation belongs to a hierarchy of infinitely many commuting flows:
\beq\label{hierarchy1}
u_{t_n}=K_n(u), \ \ \ \ n\in\NN
\eeq
i.e., such that 
\beq\label{commuteK}
[K_n(u),K_m(u)]_L:=K'_n(u)[K_m(u)]-K'_m(u)[K_n(u)]=0, \ \ \ \  n,m\in\NN,
\eeq
where
\beq\label{Frechet} 
K'_n(u)[f]=\displaystyle\lim_{\eps\to 0}{\frac{\partial K_n(u+\eps f)}{\partial \eps}}
\eeq 
is the usual Frechet derivative of $K_n(u)$ wrt $u$ in the direction $f$. 
The commuting vector fields $\{K_n\}_{n\in\NN}$ are arbitrary linear combinations, with constant coefficients, of the following 
basic symmetries $\{\sigma_n\}_{n\in\NN}$, generated by the recursion relation
\beq\label{recursion}
\ba{l}
\sigma_{n+1}=\hat R \sigma_n, \ \ \ \ \sigma_0=-i u, \ \ \ n\in\NN, \\
\hat R f:= i\left(f_x+2c u \partial^{-1}_x(u\bar f +\bar u f)\right), 
\ea
\eeq
where $\hat R$ is the recursion operator of the NLS hierarchy \cite{AKNS}. The basic symmetries used in this paper are:
\beq\label{sigma}
\ba{l}
\sigma_0=-i u, \ \ \ \ \ \ \ \sigma_2= i(u_{xx}+2 c |u|^2u),                    \\
\sigma_4=-i\Big(u_{xxxx}+2c(u^2\bar u_{xx}+2 u|u_x|^2+4 |u|^2 u_{xx}+3 u^2_x\bar u)+6c^2|u|^4u \Big), \\
\sigma_6=i\Big(u_{xxxxxx}+2c(u^2 \bar u_{xxxx}+6 |u|^2u_{xxxx}+4 u u_x\bar u_{xxx}+9 u \bar u_x u_{xxx}+     \\
15 \bar u u_x u_{xxx}+11 u |u_{xx}|^2+10 u^2_x \bar u_{xx})+10 c^2(2 u^2 |u|^2\bar u_{xx}+ 2\bar u u^2_{xx}+ \\ 
5|u_x|^2u_{xx}+5|u|^4u_{xx}+u^3\bar u^2_x+6 u |u|^2|u_x|^2+7\bar u |u|^2u^2_x)+20 c^3 u|u|^6\Big),
\ea
\eeq 
and the NLS equation (\ref{nls1}) corresponds to the flow $u_{t_2}=K_2(u)=\sigma_2(u)$.

Equivalently, the basic symmetries $\{\sigma_n\}_{n\in\NN}$ are elements of the kernel 
of the ``linearized'' $n$th flow operator $\hat M_n,~n\in\NN$, defined by
\beq\label{Mn}
\hat M_n f:=f_{t_n}-K'_n(u)[f];
\eeq   
i.e.:
\beq\label{KerM}
\hat M_n\sigma_m=0, \ \ \ n,m\in\NN. 
\eeq
Due to (\ref{commuteK}), these linearized operators commute:
\beq\label{commuteM}
\hat M_n\hat M_m=\hat M_m\hat M_n, \ \ \ n,m\in\NN.
\eeq
The linearized operators used in this paper are:
\beq
\ba{l}
\hat M_2 f:=f_{t_2}-i\Big(f_{xx}+2c(u^2\bar f+2|u|^2 f)\Big),  \\
\hat M_4 f:=f_{t_4}-\frac{i}{12}\Big[f_{xxxx}+2c(\bar u(6 u_x f_x+4 u f_{xx})+u(2 u_x{\bar f}_x+
2{\bar u}_xf_x+\\
u{\bar f}_{xx})+(6 c |u|^2u^2+3 u^2_x+4 u u_{xx}){\bar f}+(9c|u|^4+2|u_x|^2+4{\bar u}u_{xx}+ \\
2 u {\bar u}_{xx})f)\Big],         \\
\hat M_6 f:=f_{t_6}-\frac{i}{360}\Big[f_{xxxxxx}+2c(10 c u^3({\bar u}_{x}{\bar f}_{x}+{\bar u}{\bar f}_{xx})+
5(2 u^2_x{\bar f}_{xx}+\\
4{\bar u}u_{xx}f_{xx}+u_x(5{\bar u}_xf_{xx}+5u_{xx}{\bar f}_{x}+4{\bar u}_{xx}f_x+3{\bar u}f_{xxx})+
(5{\bar u}_xu_{xx}+\\
3{\bar u}u_{xxx})f_x))+u(70 c \bar u^2u_xf_x+11u_{xx}\bar f_{xx}+11 f_{xx}\bar u_{xx}+9 \bar u_xf_{xxx}+ \\
4 u_x\bar f_{xxx}+9 u_{xxx}\bar f_x+6 \bar uf_{xxxx})+u^2(5 c \bar u(6 \bar f_xu_x+6f_x\bar u_x+5\bar uf_{xx})+\\
\bar f_{xxxx})+\bar f(30 c^2|u|^4u^2+10 c u^2(3 |u_x|^2+5 \bar u u_{xx})+10 c u^3\bar u_{xx}+\\
5(2 u^2_{xx}+3 u_xu_{xxx})+u(70 c\bar u u^2_x+6 u_{xxxx}))+f(40 c^2 |u|^6+35 c \bar u^2u^2_x+\\
11 |u_{xx}|^2+15 c u^2(\bar u^2_x+2\bar u\bar u_{xx})+9\bar u_xu_{xxx}+4 u_x\bar u_{xxx}+6 \bar uu_{xxxx}+\\
2 u(5c\bar u(6|u_x|^2+5 \bar u u_{xx})+\bar u_{xxxx})))\Big].
\ea
\eeq
At last, if $c=0$, equations (\ref{nls1}) and (\ref{recursion}) lead to the linear Schr\"odinger (LS) equation 
\beq\label{LS}
iu_t+u_{xx}=0
\eeq
and to its (trivial) symmetries $(-i^{n+1}\partial^n_xu)$. 

The paper is organized as follows. In \S 2 we construct the multiscale expansion, in the small $\eps$ regime, of 
a generic solution of (\ref{dnlsF}), establishing, in particular, that the leading term of such expansion evolves 
wrt the infinitely many ``even'' time variables $t_{2k}:=\eps^{2(k-1)}t,~k\in\NN_+$ according to the even flows of the NLS 
hierarchy. In \S 3, after summarizing the DP test and after showing how to use the main output of this test 
to construct infinitely many approximate symmetries of the original P$\Delta$E through novel and simple formulas, 
we apply the DP test to the P$\Delta$E (\ref{dnlsF}), isolating the constraints on the coefficients ${a_j}'s,~j=1,..,10$ 
allowing one to pass the test at time scales of $O(\eps^{-2})$ and of $O(\eps^{-4})$, in both scenarios of C- and S- integrability. 
In \S 4 we summarize the results of the paper and we discuss the research perspectives opened by this work. In the 
Appendix (\S 5) we display the long outputs of the DP test, obtained using the algebraic manipulation program of 
Mathematica. 

\section{Multiscale expansion in the small lattice spacing regime}
If the lattice spacing $\eps$ is small: $0<\eps<<1$, as consequence of the invariance of (\ref{dnlsF}) under the 
transformation $\psi_{n\pm 1}\to \psi_{n\mp 1}$ and of the well-known formula
\beq\label{cont}
f_{n\pm 1}=\sum\limits_{k=0}^{\infty}\frac{(\pm 1)^k}{k!}\eps^k \partial^k_x f,
\eeq 
only even $x$-derivatives appear at all (even) orders in $\eps$, implying that also the asymptotic expansion of $\psi_n$ 
contains only even powers of $\epsilon$. Consequently, to eliminate the secularities appearing at all even orders in $\eps$, 
the coefficients of such expansion must depend on infinitely many ``even'' slow times \cite{DMS}:
\beq\label{multitime}
\vec t=(t_2,t_4,t_6,\dots), \ \ \ t_{2k}:=\eps^{2(k-1)}t, \ k\in\NN_+ ,
\eeq 
implying that
\beq\label{multitime_derivatives}
\partial_t \to \partial_{t_2}+\eps^2\partial_{t_4}+\eps^4\partial_{t_6}+\dots .
\eeq    
Therefore we are lead to the following ansatz for the asymptotic expansion of the ``generic'' solution of (\ref{dnlsF})
\beq\label{expansion}
\psi_n(t)=\sum\limits_{k=0}^{\infty}\eps^{2k}u^{(2k+1)}(x,\vec t), ~~~u^{(1)}(x,\vec t)=u(x,\vec t).
\eeq
Plugging (\ref{cont}), (\ref{multitime_derivatives}) and (\ref{expansion}) into equation (\ref{dnlsF}) and 
equating to zero the coefficients of all powers in $\eps$, we obtain the following results.

At the leading $O(1)$, we obtain the NLS equation for the leading term $u^{(1)}=u$ wrt the first time $t_2=t$:
\beq
\ba{l}
u_{t_2}=K_2(u), \\
K_2(u):=\sigma_2(u)=i(u_{xx}+2c |u|^2u), ~~c=\sum\limits_{j=1}^{10}a_j.
\ea
\eeq
As usual in perturbation theory, at the next relevant order ($O(\eps^2)$ in our case), 
the ``linearization'' $\hat M_2 u^{(3)}$ of $(u_{t_2}-K_2(u))$ appears, together with the linear term 
($u_{t_4}-(i/12)u_{xxxx}$), coming from the linear part of (\ref{dnlsF}), and with a nonlinear term $G_5$, coming 
from the nonlinear part of (\ref{dnlsF}):
\beq\label{coeff1a}
\hat M_2u^{(3)}=-\left(u_{t_4}-i\frac{2}{4!}u_{xxxx}\right) + G_5,
\eeq
where 
\beq
\ba{l}
G_5=i\left(s_1 {u}^2\bar u_{xx}+s_2 |u|^2u_{xx}+s_3 u |u_x|^2+ 
s_4 \bar u {u^2_x}\right), \\
s_1=a_3+a_4+a_5+a_8+a_9+a_{10},~~\\ s_2=a_2+a_4+a_5+ 2(a_6+ a_7+ a_8+a_9+a_{10}), \\s_3=2(a_4-a_5+2a_8-2 a_9),~~
s_4=2(a_6-a_7+a_8+a_9-a_{10}). 
\ea
\eeq
Concentrating on the linear terms in round bracket, we observe that $u_{t_4}\in$ Ker$\hat M_2$ and ($-(i/12)u_{xxxx}$) 
is the linear 
part of the symmetry $(2/4!)\sigma_4(u)\in$ Ker$\hat M_2$. Therefore, adding and subtracting the symmetry $(2/4!)\sigma_4$, 
equation (\ref{coeff1a}) is conveniently rearranged in the following way, isolating the resonant terms in round bracket:
\beq\label{coeff1b}
\ba{l}
\hat M_2u^{(3)}=-\left(u_{t_4}+\frac{2}{4!}\sigma_4(u)\right)+g_5, 
\ea
\eeq
where
\beq\label{g5}
\ba{l}
g_5:=i(c_1 |u|^4u+c_2 \bar u {u_x}^2+c_3 u|u_x|^2+c_4 |u|^2u_{xx}+c_5 u^2\bar u_{xx}) 
\ea
\eeq
and 
\beq\label{cj}
\ba{l}
c_1=-\frac{1}{2}c^2,\\
c_2=-\frac{1}{2}(a_1+a_2+a_3+a_4+a_5-3 a_6+5 a_7-3 a_8-3 a_9+5 a_{10}), \\
c_3=-\frac{1}{3}(a_1+a_2+a_3-5 a_4+7 a_5+a_6+a_7-11 a_8+13 a_9+a_{10}), \\
c_4=-\frac{1}{3}[2 a_1-a_2+2 a_3-a_4-a_5-4(a_6+a_7+a_8+a_9+a_{10})], \\
c_5=-\frac{1}{6}[a_1 + a_2 + a_6 + a_7 - 5 (a_3 +a_4+ a_5 +a_8 +a_9+ a_{10})]. 
\ea
\eeq
To eliminate the secularity in bracket, we are forced to choose
\beq
u_{t_4}=K_4(u):=-\frac{2}{4!}\sigma_4(u),
\eeq 
so that (\ref{coeff1b}) finally becomes the following secularity free equation for the first correction $u^{(3)}$:
\beq\label{coeff1c}
\ba{l}
\hat M_2u^{(3)}=g_5.
\ea
\eeq
This procedure iterates without essential differences at all orders. The terms $\hat M_2 u^{(3)}$ and 
$(u_{t_4}-K_4(u))$ in (\ref{coeff1b}) generate, at $O(\eps^4)$, the terms $\hat M_2 u^{(5)}$ and $\hat M_4 u^{(3)}$ 
respectively, while the new linear term $(u_{t_6}-i(2/6!)u_{xxxxxx})$  
is rearranged again into the secular factor $(u_{t_6}-(2/6!)\sigma_6)$ that must be set to zero, to avoid secularities. 
Since, at $O(\eps^{2k})$, we produce the linear term  
$\left(u_{t_{2k}}-i\frac{2}{(2k)!}\partial^{2k}_xu \right)$, one infers, in analogy with \cite{DMS}, that 
{\it $u$ evolves wrt to the higher times according to the even flows of the NLS hierarchy as follows:}
\beq\label{higherflows}
u_{t_{2k}}=K_{2k}(u):=(-)^{k+1}\frac{2}{(2k)!}\sigma_{2k}(u),~~k\in\NN_+,
\eeq
and one is left with the following triangular set of equations \cite{DP,Degasperis}
\beq\label{M_equs}
\ba{cr}
O(\eps^2): & \hat M_2 u^{(3)}=g_5, \\
O(\eps^4): & \hat M_2 u^{(5)}+\hat M_4 u^{(3)}=g_7, \\
O(\eps^6): & \hat M_2 u^{(7)}+\hat M_4 u^{(5)}+\hat M_6 u^{(3)}=g_9, \\
\vdots &                 \vdots \\
O(\eps^{2k}): & \hat M_2 u^{(2k+1)}+\hat M_4 u^{(2k-1)}+\hdots +\hat M_{2k} u^{(3)}=g_{2k+3}, \\ 
\vdots &                 \vdots \\ 
\ea
\eeq
where, f.i., the expression of $g_7$ is presented in formula (\ref{g7}) of the Appendix. 
It remains to remark, following \cite{DP,Degasperis}, that the symmetries $\{\sigma_n\}$ and the expressions 
in (\ref{M_equs}) bring naturally to the definition of the following vector spaces. \\
{\bf Definitions} Let ${\cal P}_n$ be the vector space of all differential polynomials in the functions $(\partial^j_xu^{(k)})$ and 
$(\partial^j_x\overline{u^{(k)}})$ of order $n$, possessing the gauge symmetry of first kind, where:
\beq
\mbox{order }(\partial^j_xu^{(k)})=\mbox{order }(\partial^j_x\overline{u^{(k)}})=j+k
\eeq
(so that order$(\partial^j_xu)=j+1$, since $u=u^{(1)}$),  
and let ${\cal P}_n(m)$ be the subspace of ${\cal P}_n$ of all differential polynomials in the functions $(\partial^j_xu^{(k)})$ and 
$(\partial^j_x\overline{u^{(k)}})$, of order $n$, possessing the gauge symmetry of first kind and such that $k\le m$.      \\
Is is easy to see that, for instance, $\sigma_n,K_n\in{\cal P}_{n+1}(1)$, $g_5\in{\cal P}_5(1)$ and $g_7\in{\cal P}_7(3)$ (see 
(\ref{g7})).

\section{Applying the DP integrability test}

Suppose we generate, from the model to be tested, a NLS-type multiscale expansion (as in our example); then we have 
the following scenarios. 
If such a model is S-integrable (C-integrable), \\
1) the leading term $u$ of the asymptotic expansion evolves,  
with respect to the slow times $t_n$, according to the NLS (LS) hierarchy \cite{DMS}; \\
2) there exist elements 
$f^{(m)}_n\in{\cal P}_{n+m}$ such that the following equations hold \cite{DP,Degasperis}:
\beq\label{Mu=f}
\hat M_n u^{(m)}=f^{(m)}_n\in{\cal P}_{m+n},~~m,n\in\NN_+ ,
\eeq 
implying, due to (\ref{commuteM}), the compatibility conditions
\beq\label{comp}
\hat M_n f^{(j)}_m=\hat M_m f^{(j)}_n,~~m,n,j\in\NN_+.
\eeq
Therefore equations (\ref{comp}) are necessary conditions to be satisfied, in cascade, for the model 
under investigation to be S -(C -) integrable; they are also sufficient to guaranty the asymptotic 
character of the expansion. If equations (\ref{comp}) are satisfied only up to 
a certain order, the model under investigation is not integrable, being nevertheless ``asymptotically integrable up to that 
order'' \cite{DP,Degasperis}.

\subsection{The DP test and approximate symmetries}

Equations (\ref{Mu=f}),(\ref{comp}), the basic formulae of the DP test, have been derived in 
\cite{DP,Degasperis} as a consequence of the existence of a Lax pair for the starting integrable model. 
It follows that, if the conditions (\ref{Mu=f}),(\ref{comp}) are satisfied up to a certain order, the equation 
under scrutiny admits an approximate Lax pair up to that order. 

In this subsection we show how to derive the conditions (\ref{Mu=f}), (\ref{comp}) from the existence of infinitely-many 
symmetries of the starting integrable model. This derivation allows one to establish the important  
relations (to the best of our knowledge so far unknown) between the functions $f^{(m)}_n\in{\cal P}_{m+n}$ of the DP test and the   
symmetries of the starting model. We concentrate our attention on the case of difference equations, but our considerations 
have general validity.

Let ${\psi_n}_{t_2}={\cal K}_2(\psi_n)$ be an integrable model, say, the AL equation (\ref{AL}), and let 
${\psi_n}_{t_{2m}}={\cal K}_{2m}(\psi_n),~m>2$, be one of its infinitely-many higher order commuting flows (symmetries), 
reducing, in the continuous limit, to the higher commuting flow $u_{t_{2m}}=K_{2m}(u)$ of NLS. 

On one hand, from equations (\ref{multitime}),(\ref{expansion}), we have that 
\beq\label{symm1}
{\psi_n}_{t_{2m}}=u_{t_{2m}}+\eps^2(u_{t_{2(m+1)}}+u^{(3)}_{t_{2m}})+\dots =
\sum\limits_{k\ge 0}\eps^{2k}\left(\sum\limits_{j=m}^{m+k}u^{(2(m+k-j)+1)}_{t_{2j}}\right),
\eeq 
where $u_{t_{2m}}=K_{2m}(u)$ (from (\ref{higherflows})) and $u^{(2(m+k-j)+1)}_{t_{2j}}=K'_{2j}[u^{(2(m+k-j)+1)}]+f^{(2(m+k-j)+1)}_{2j}$, 
for some functions $f^{(2(m+k-j)+1)}_{2j}$ to be specified. On the other hand:
\beq\label{symm2}
{\cal K}_{2m}(\psi_n)=K_{2m}(u)+\eps^2 K^{(2)}_{2m}+\dots=K_{2m}(u)+\sum\limits_{k\ge 1}\eps^{2k}K^{(2k)}_{2m},
\eeq  
where $K^{(2k)}_{2m}\in {\cal P}_{2(m+k)+1}$. Equating equations (\ref{symm1}) and (\ref{symm2}), we infer that 
$f^{(m)}_n\in{\cal P}_{m+n},~~m,n\in\NN_+$ (the basic formula (\ref{Mu=f}) of the DP test), and we also construct 
the asymptotic expansion of the generic higher order symmetry
\beq\label{symm3}
\ba{l}
{\cal K}_{2m}(\psi_n)=K_{2m}(u)+\eps^2\left(K_{2(m+1)}(u)+K'_{2m}[u^{(3)}]+f^{(3)}_{2m}\right)+\dots =\\
\sum\limits_{k\ge 0}\eps^{2k}\left(\sum\limits_{j=m}^{m+k-1}\left(K'_{2j}[u^{(2(m+k-j)+1}]+f^{(2(m+k-j)+1)}_{2j}\right)+
K_{2(m+k)}(u)\right)
\ea
\eeq 
in terms of the NLS higher order symmetries, of their Frechet derivatives in the direction of the corrections $u^{(j)},~j>1$ 
of the leading term $u$ of the expansion (\ref{expansion}), and of the output functions $f^{(m)}_n\in {\cal P}_{m+n}$ of the 
DP test. 

Therefore, if $f^{(2k+1)}_{2n}\in {\cal P}_{2(k+n)+1}$ esists, but $f^{(2k+3)}_{2n}\in {\cal P}_{2(k+n)+3}$ does not, 
$\forall n\in\NN_+$, it follows that: \\
i) the solution $u^{(2k+1)}$ of (\ref{Mu=f}) is uniformely bounded and the expansion (\ref{expansion}) is asymptotic 
up to the $O(\eps^{2k})$; therefore the P$\Delta$E under scrutiny approximates well its continuous limit, with an error 
of $O(\eps^2)$, for time scales up to the $O(\eps^{-2k})$. \\
ii) The P$\Delta$E possesses infinitely-many ``approximate'' generalized symmetries in the form (\ref{symm3})  
up to the $O(\eps^{2k})$; therefore it is integrable up 
to that order. We remark that, due to the Hamiltonian theory of integrable systems \cite{Magri}-\cite{SF}, it is also 
possible to associate with the P$\Delta$E infinitely-many ``approximate'' constants of motion in involution, a very useful 
information in a any numerical check.

\subsection{C - and S - integrability at $O(\eps^2)$}

In our example, the first of equations (\ref{M_equs}) is already in the form (\ref{Mu=f}), with $g_5=f^{(3)}_2\in{\cal P}_5(1)$. 
Assuming now that $\hat M_4 u^{(3)}=f^{(3)}_4$, we arrive at the consistency 
\beq\label{cond1}
\hat M_4f^{(3)}_2=\hat M_2f^{(3)}_4,
\eeq 
that must be viewed as an equation for the unknown $f^{(3)}_4$. Since $g_5=f^{(3)}_2\in{\cal P}_{5}(1)$, it follows that one 
must look for 
$f^{(3)}_4\in{\cal P}_{7}(1)$. The calculation, plain but lengthy, has been performed using the algebraic manipulation 
program of Mathematica, and gives the following result.

\noindent
{\bf Lemma 1}. Equation (\ref{cond1}) admits a unique solution $f^{(3)}_4\in{\cal P}_{7}(1)$ (presented in formula (\ref{f43}) 
of the Appendix) iff the coefficients $a_j$'s appearing in (\ref{dnlsF}) satisfy the following quadratic constraint
\beq\label{constr1}
(a_1-3a_3-2a_4-6a_5-5 a_6 +3a_7-5a_8-13a_9-a_{10})(\sum\limits_{j=1}^{10}a_j)=0.
\eeq   
Once $f^{(3)}_4$ is constructed, $f^{(5)}_2\in{\cal P}_7(3)$ ($f^{(5)}_2=\hat M_2u^{(5)}$) is found from the second of 
equations (\ref{M_equs}): 
\beq\label{f25}
f^{(5)}_2=g_7-f^{(3)}_4 
\eeq
and is presented in formula (\ref{f25}) of the Appendix. 

We first notice the nice factorization of the quadratic constraint (\ref{constr1}) into two linear contraints:
\beq\label{Cint_1}
c=\sum\limits_{j=1}^{10}a_j=0,
\eeq
\beq\label{Sint_1}
a_1-3a_3-2a_4-6a_5-5 a_6+3a_7-5a_8-13a_9-a_{10}=0.
\eeq
Therefore we have the following two different scenarios. 
\vskip 5pt
\noindent
1) If the first constraint (\ref{Cint_1}) is satisfied by the coefficients $a_j$'s, the continuous limits of dNLS (\ref{dnlsF}) 
is the linear Schr\"odinger 
equation. It follows that, in this case, equation (\ref{dnlsF}) is ``asymptotically C-integrable'' at $O(\eps^2)$ and 
one expects that, 
for generic initial data and at time scales of $O(\eps^{-2})$, the dynamics according to (\ref{dnlsF}),(\ref{Cint_1}) 
be well approximated by the dynamics according to the linear Schr\"odinger equation (\ref{LS}) with an error of $O(\eps^2)$.

In particular, the dNLS (\ref{dnlsF}),(\ref{OJE}) is ``asymptotically C-integrable'' at $O(\eps^2)$ iff:
\beq
a_1+4 a_3+ 3 a_6=0.
\eeq 
\vskip 5pt
\noindent
2) If, instead, the second constraint (\ref{Sint_1}) is satisfied by the coefficients $a_j$'s, the dNLS equation (\ref{dnlsF})  
is ``asymptotically S-integrable'' at $O(\eps^2)$  and one expects that, 
for generic initial data and at time scales of $O(\eps^{-2})$, the dynamics according to the dNLS equation 
(\ref{dnlsF}), (\ref{Sint_1}) approximates well the dynamics according to the NLS equation (\ref{nls1}),(\ref{c}) 
with an error of $O(\eps^2)$.

In particular, i) the dNLS (\ref{dnlsF}),(\ref{Peli}) is ``asymptotically S-integrable'' at $O(\eps^2)$ iff the 
following additional constraint is satisfied:  
\beq\label{Sint_1_Peli}
a_3+2 a_8=0;
\eeq
ii) the dNLS (\ref{dnlsF}),(\ref{OJE}) is ``asymptotically S-integrable'' at $O(\eps^2)$ iff the 
following additional constraint is satisfied:
\beq\label{Sint_1_OJE}
a_1-8 a_3-9 a_6=0,
\eeq 
while the dNLS (\ref{dnlsF}),(\ref{CKKS}) is not ``asymptotically S-integrable'' at $O(\eps^2)$ (therefore it is not integrable). 
 
In addition, since the dNLS equation (\ref{dnlsF}) is the linear combination of ten  
different discretizations of NLS, it is immediate to check if some of these ten discretizations satisfy the 
constraint (\ref{Sint_1}). 
Calling dNLS$_k$ the single discretization of NLS obtained choosing in (\ref{dnlsF}) $a_j=a_k\delta_{jk},~j=1,\dots,10$, 
it is straightforward to see (since the coefficient $a_2$ is the only one absent in (\ref{Sint_1})) that only the dNLS$_2$ 
equation (coinciding with the AL equation (\ref{AL})) satisfies the constraint (\ref{Sint_1}) (as it has to be, being an 
integrable system). All the other dNLS$_k,~k\ne 2$ equations, including the dNLS$_1$ equation (\ref{dNLS1}), do not satisfy the 
constraint (\ref{Sint_1}); therefore they are not ``asymptotically S-integrable'' at $O(\eps^2)$ (consequently, they are 
not integrable) and, for generic initial data and at time scales of $O(\eps^{-2})$, their dynamics are expected to be  
quite different from that of NLS (\ref{nls1}),(\ref{c}), presumably exhibiting numerical evidence of nonintegrability 
and/or chaos. 
   
We finally infer that the discretizations (\ref{dnlsF}),(\ref{Peli}) and (\ref{dnlsF}),(\ref{OJE}) satisfying 
respectively the constraints (\ref{Sint_1_Peli}) and (\ref{Sint_1_OJE}), the AL equation and any other dNLS equation 
(\ref{dnlsF}) satisfying the constraint (\ref{Sint_1}) are all close to NLS (once the free coefficients of each model are 
normalized to satisfy (\ref{c})) and are all close together at time scales of $O(\eps^{-2})$, in the sense 
mentioned in the introduction.

It is interesting now to push the integrability test to the next order. Due to the above factorization of 
the constraint (\ref{constr1}), the test bifurcates and, in the next two subsections, we explore both cases. 
Before doing that, we observe that, given $f_2^{(3)}\in{\cal P}_{5}(1)$ and assuning that the constraint (\ref{constr1}) be satified, 
equation $\hat M_6 f_2^{(3)}=\hat M_2 f_6^{(3)}$ 
admits a unique solution $f_6^{(3)}=\hat M_6 u^{(3)}\in{\cal P}_{9}(1)$, presented in formula (\ref{f63}) of the Appendix, and no 
additional constraint appears in this derivation, as predicted by the DP test.

\subsection{C - integrability at $O(\eps^4)$}

Let us assume that the constraint (\ref{Cint_1}) be satified.  For the construction of 
$f^{(5)}_4=\hat M_4 u^{(5)}\in{\cal P}_{9}(3)$ from the equation 
\beq\label{cond3}
\hat M_4 f^{(5)}_2=\hat M_2 f^{(5)}_4
\eeq
we have the following result. \\
{\bf Lemma 2} Equation (\ref{cond3}) admits a unique solution $f^{(5)}_4\in{\cal P}_{9}(3)$, presented in formula (\ref{f45}) of 
the Appendix, iff the 
coefficients $a_j$'s satisfy the four linear constraints (\ref{Cint_3}), 
defining a 6-parameter family (but one of these parameters can always be rescaled away) of dNLS equations (\ref{dnlsF}) 
``asymptotically C-integrable'' at $O(\eps^4)$. Therefore one  
expects that, for generic initial data and at time scales of $O(\eps^{-4})$, the dynamics according to (\ref{dnlsF}),(\ref{Cint_3}) 
well approximate the dynamics according to the linear Schr\"odinger equation (\ref{LS}) with an error of $O(\eps^2)$.

For instance, the discretization (\ref{dnlsF}),(\ref{OJE}) satisfies the constraints (\ref{Cint_3}) iff $a_6=-a_3=a_1$.  

The 6-parameter family of dNLS equations (\ref{dnlsF}),(\ref{Cint_3}) (or at least some particular case of it), 
being C-integrable at such a high order, is a natural candidate to be a C-integrable discrete system.   

\subsection{S - integrability at $O(\eps^4)$}

Let us assume that the constraint (\ref{Sint_1}) be satified.  For the construction of a unique  
$f^{(5)}_4=\hat M_4 u^{(5)}\in{\cal P}_{9}(3)$ from equation (\ref{cond3}), we have the following result.

\vskip 5pt
\noindent
{\bf Lemma 3}. If the constraint (\ref{Sint_1}) is satified, equation (\ref{cond3}) admits a unique 
solution $f^{(5)}_4\in{\cal P}_{9}(3)$, presented in formula (\ref{f45}) of the Appendix, iff the following five quadratic 
constraints are satisfied: 
\beq\label{S4}
Q_j=0,~j=1,\dots,5,
\eeq
where the $Q_j$'s are the following quadratic forms in the 9 variables \\$a_2, \dots, a_{10}$:
\beq\label{defQ1}
\ba{l}
Q_1 = -4 a^2_{10} + a_{10} a_2 + 2 a_{10} a_3 - a_2 a_3 + 2 a^2_3 - a_{10} a_4 - 2 a_2 a_4 + 
 a_3 a_4 + \\ 
3 a_{10} a_5 - 2 a_2 a_5 - 3 a_3 a_5 - 8 a_4 a_5 - 8 a^2_5 + 18 a_{10} a_6 + 6 a_3 a_6 - 6 a_{10} a_7 + \\
4 a_2 a_7 + 6 a_3 a_7 + 4 a_4 a_7 + 20 a_5 a_7 + 24 a_6 a_7 - 8 a^2_7 + 12 a_{10} a_8 - 3 a_2 a_8 + \\
6 a_3 a_8 + 3 a_4 a_8 - 9 a_5 a_8 - 6 a_6 a_8 + 18 a_7 a_8 + 20 a_{10} a_9 - 3 a_2 a_9 - 2 a_3 a_9 - \\
13 a_4 a_9 - 25 a_5 a_9 - 6 a_6 a_9 + 50 a_7 a_9 - 24 a_8 a_9 - 24 a^2_9, 
\ea
\eeq
\beq\label{defQ2}
\ba{l}
Q_2=14 a^2_{10} + 6 a_{10} a_2 + 44 a_{10} a_3 + 4 a_2 a_3 + 26 a^2_3 + 36 a_{10} a_4 + 5 a_2 a_4 + \\
40 a_3 a_4 + 17 a^2_4 + 72 a_{10} a_5 + 7 a_2 a_5 + 88 a_3 a_5 + 68 a_4 a_5 + 75 a^2_5 + \\
64 a_{10} a_6 + 
10 a_2 a_6 + 72 a_3 a_6 + 60 a_4 a_6 + 128 a_5 a_6 + 60 a^2_6 - 24 a_{10} a_7 - \\
2 a_2 a_7 - 24 a_3 a_7 - 
16 a_4 a_7 - 44 a_5 a_7 - 32 a_6 a_7 + 4 a^2_7 + 64 a_{10} a_8 + 8 a_2 a_8 + \\
60 a_3 a_8 + 54 a_4 a_8 + 
106 a_5 a_8 + 100 a_6 a_8 - 20 a_7 a_8 + 42 a^2_8 + 168 a_{10} a_9 + \\
28 a_2 a_9 + 220 a_3 a_9 + 
170 a_4 a_9 + 382 a_5 a_9 + 332 a_6 a_9 - 108 a_7 a_9 + \\
284 a_8 a_9 + 466 a^2_9,
\ea
\eeq                                                                           
\beq\label{defQ3}
\ba{l}
Q_3=20 a^2_{10} + 15 a_{10} a_2 + 38 a_{10} a_3 - 5 a_2 a_3 - 22 a^2_3 + 39 a_{10} a_4 - a_2 a_4 - \\
23 a_3 a_4 - a^2_4 + 63 a_{10} a_5 - 11 a_2 a_5 - 83 a_3 a_5 - 52 a_4 a_5 - 75 a^2_5 + \\
70 a_{10} a_6 - 14 a_2 a_6 - 78 a_3 a_6 - 48 a_4 a_6 - 148 a_5 a_6 - 84 a^2_6 - 18 a_{10} a_7 + \\
10 a_2 a_7 + 42 a_3 a_7 + 32 a_4 a_7 + 76 a_5 a_7 + 88 a_6 a_7 - 20 a^2_7 + 88 a_{10} a_8 - \\
7 a_2 a_8 - 54 a_3 a_8 - 15 a_4 a_8 - 119 a_5 a_8 - 134 a_6 a_8 + 82 a_7 a_8 - 36 a^2_8 + \\
72 a_{10} a_9 - 59 a_2 a_9 - 302 a_3 a_9 - 211 a_4 a_9 - 539 a_5 a_9 - 526 a_6 a_9 + 234 a_7 a_9 - \\
472 a_8 a_9 - 788 a^2_9, 
\ea
\eeq                                                                          
\beq\label{defQ4}
\ba{l}
Q_4=-32 a^2_{10} - 24 a_{10} a_2 - 56 a_{10} a_3 + 6 a_2 a_3 + 36 a^2_3 - 70 a_{10} a_4 - a_2 a_4 + \\
30 a_3 a_4 + a^2_4 - 114 a_{10} a_5 + a_2 a_5 + 78 a_3 a_5 + 20 a_4 a_5 + 27 a^2_5 - 120 a_{10} a_6 + \\
14 a_2 a_6 + 88 a_3 a_6 + 48 a_4 a_6 + 84 a_5 a_6 + 84 a^2_6 + 24 a_{10} a_7 - 14 a_2 a_7 - 64 a_3 a_7 - \\
52 a_4 a_7 - 80 a_5 a_7 - 112 a_6 a_7 + 28 a^2_7 - 164 a_{10} a_8 + 2 a_2 a_8 + 48 a_3 a_8 + 12 a_4 a_8 + \\ 
16 a_5 a_8 + 124 a_6 a_8 - 116 a_7 a_8 + 36 a^2_8 - 220 a_{10} a_9 + 22 a_2 a_9 + 208 a_3 a_9 + 96 a_4 a_9 + \\
196 a_5 a_9 + 292 a_6 a_9 - 204 a_7 a_9 + 176 a_8 a_9 + 300 a^2_9,
\ea
\eeq                                   
\beq\label{defQ5}
\ba{l}
Q_5=4 a^2_{10} + 3 a_{10} a_2 - 2 a_{10} a_3 - a_2 a_3 - 14 a^2_3 + 3 a_{10} a_4 - a_2 a_4 - 19 a_3 a_4 - \\
5 a^2_4 - 5 a_{10} a_5 - 3 a_2 a_5 - 47 a_3 a_5 - 36 a_4 a_5 - 39 a^2_5 - 2 a_{10} a_6 - 6 a_2 a_6 - \\
38 a_3 a_6 - 32 a_4 a_6 - 68 a_5 a_6 - 36 a^2_6 + 6 a_{10} a_7 + 2 a_2 a_7 + 18 a_3 a_7 + 16 a_4 a_7 + \\
28 a_5 a_7 + 24 a_6 a_7 - 4 a^2_7 + 8 a_{10} a_8 - 3 a_2 a_8 - 30 a_3 a_8 - 19 a_4 a_8 - 59 a_5 a_8 - \\
62 a_6 a_8 + 26 a_7 a_8 - 20 a^2_8 - 40 a_{10} a_9 - 23 a_2 a_9 - 150 a_3 a_9 - 119 a_4 a_9 - 255 a_5 a_9 - \\
230 a_6 a_9 + 82 a_7 a_9 - 216 a_8 a_9 - 356 a^2_9. 
\ea
\eeq  
The five homogeneous quadratic constraints (\ref{S4}),(\ref{defQ1})-(\ref{defQ5}) for nine unknowns, 
characterizing the intersection 
of 5 quadrics in the real projective space of dimension $8$, define, in principle, a 4 - parameter 
family of solutions (but one of these parameters can always be rescaled away) whose parametrization does not 
appear to be expressible, in general, in terms of elementary functions. 
The corresponding dNLS equation (\ref{dnlsF}) is asymptotically S-integrable at $O(\eps^4)$ and should well 
approximate the NLS equation for times up to the $O(\eps^{-4})$.  

We observe that, in all these quadratic constraints, $a_2$ is the only coefficient 
appearing always multiplied by other coefficients (the term $(a_2)^2$ is absent); therefore the choice 
\beq
a_j=a_2 \delta_{j2},~j=1,\dots,10,
\eeq 
corresponding to the AL equation (\ref{AL}), satisfies all constraints, as it 
has to be. Other less trivial explicit solutions of (\ref{Sint_1}),(\ref{S4}),(\ref{defQ1})-(\ref{defQ5}) can also be 
constructed, corresponding to the case in which all quadrics degenerate into pairs of hyperplanes. 
Here we display the following two examples:
\beq
\ba{l}
a_1=-4 a_6,~~a_2=\frac{4 a_6}{3},~~a_3=4 a_6,~~a_4=0,\\
a_5=-4 a_6,~~a_7=-a_6,~~a_8=a_9=a_{10}=0,
\ea
\eeq
\beq
\ba{l}
a_1=-24 a9,~~a_2=a_3=0,~~a_4=a_5=-8 a_9, \\
a_6=10 a_9,~~a_7=-2 a_9,~~a_8=-7 a_9,~~a_{10}=6 a_9, 
\ea
\eeq
corresponding, respectively, to the dNLS equations (\ref{dNLS_S_O4_A}) and (\ref{dNLS_S_O4_B}) presented in the Introduction,  
``asymptotically S-integrable'' at $O(\eps^4)$. Therefore one  
expects that, for generic initial data and at time scales of $O(\eps^{-4})$, the dynamics according to 
equations (\ref{dNLS_S_O4_A}) and (\ref{dNLS_S_O4_B})  
are good approximations of the dynamics according to the NLS equation (\ref{nls1}), with an error of $O(\eps^2)$,  
at time scales of $O(\eps^{-4})$.  Of course, these distinguished equations, passing the test at such a high order, 
are also good candidates to be S-integrable difference equations. 

We finally observe that there is no choice of parameters for which the dNLS equations (\ref{dnlsF}),(\ref{Peli}) and 
(\ref{dnlsF}),(\ref{OJE}) satisfy the above constraints; therefore these two models are not S-integrable at this order 
(they are not S-integrable at all) and do not approximate well NLS at time scales of $O(\eps^{-4})$.

\section{Summary of the results and future perspectives}
In this paper we have proposed an algorithmic procedure allowing one: i) to study the distance between an integrable PDE and 
any P$\Delta$E discretizing it, in the small lattice spacing $\eps$ regime; ii) to test the 
(asymptotic) integrability properties of such a P$\Delta$E, and iii) to construct infinitely many (approximate) symmetries 
and conserved quantities for it. This method should provide, in particular,  
useful and concrete informations on how good is a numerical scheme used to integrate a given integrable PDE.

The procedure we have proposed, illustrated on the basic prototype example of the nonlinear Schr\"odinger equation (\ref{nls1}) 
and of its discretization (\ref{dnlsF}), consists in the following three steps: i) the construction of 
the multiscale expansion of a 
generic solution of the dNLS (\ref{dnlsF}) at all orders in $\eps$, following \cite{DMS}; ii) the  
application, to such expansion, of the DP integrability test \cite{DP,Degasperis}; iii) the use the main output 
of such test to construct infinitely many approximate symmetries of the dNLS equation (\ref{dnlsF}), through  
novel formulas presented in this paper. 

This approach allows one to study the distance between the integrable PDE and 
any P$\Delta$E discretizing it. Suppose, for instance, that the asymptotic 
expansion we construct reads $\psi=u+O(\eps^{\alpha}),~\alpha>0$, where $\psi$ is a generic solution of the dNLS (\ref{dnlsF}) and $u$ 
is the corresponding solution of (\ref{nls1}), then if the DP test is passed at $O(\eps^{\beta})$, we conclude that:  
i) the dynamics according to the NLS equation (\ref{nls1}) is well approximated (with an error of $O(\eps^{\alpha})$) by the dynamics 
according to its discretization (\ref{dnlsF}), for time scales of $O(t^{-\beta})$; ii) the dNLS equation is 
asymptotically integrable up to that order, constructing its infinitely many approximate symmetries and constants of motion 
in involution.  
On the contrary, if the DP test is not passed at that order, the dNLS equation is not integrable and one should expect, 
at the corresponding time scale, numerical evidence of nonintegrability and/or chaos.

We have carried the above procedure up to the $O(\eps^4)$ and we have been able to isolate the constraints on the coefficients of 
the dNLS equation (\ref{dnlsF}) allowing one to pass the test at that order, in both scenarios of S- and C- integrability. 

Numerical experiments to test such theoretical findings are presently under investigation; preliminary results seem 
to confirm the theoretical predictions contained in this paper \cite{AS}.     
   
With the same methodology and goals, we are presently investigating families of discretizations of 
the Korteweg-de Vries and Burgers equations \cite{ASS}, other two basic integrable models of natural phenomena. Of 
course we also plan to investigate discretizations of integrable PDEs in which also the time variable is discretized.

\section{Appendix}
\begin{footnotesize}
In this Appendix we display, for completeness, the long outputs of the DP test, obtained using the algebraic manipulation program 
of Mathematica. 

The differential polynomial $g_7$ in (\ref{M_equs}) reads: 
\beq\label{g7}
\ba{l}
g_7= i\Big(l_{1} u |u|^6 +l_{2} |u|^4 u^{(3)}+l_{3} \bar u{u^{(3)}}^2+ l_{4} u^2 |u|^2\bar u^{(3)}+ l_{5} u |u^{(3)}|^2+ \\ 
l_{6}\bar u |u|^2{u^2_x}+ l_{7} {u^2_x} \bar u^{(3)}+l_{8} u |u|^2|u_x|^2+l_{9}|u_x|^2 u^{(3)}+l_{10} u^3 \bar {u}^2_x +\\
l_{11} \bar u u_x u^{(3)}_x+l_{12} u \bar u_x u^{(3)}_x+l_{13} u u_x \bar u^{(3)}_x +l_{14}|u|^4u_{xx}+l_{15}\bar u u_{xx}u^{(3)}+\\
l_{16} u u_{xx}\bar u^{(3)}+l_{17} |u_x|^2u_{xx}+l_{18}\bar u {u^2_{xx}}+l_{19} u^2 |u|^2 \bar u_{xx}+l_{20} u\bar u_{xx}u^{(3)}+ \\
l_{21} u^2_x\bar u_{xx}+l_{22} u |u_{xx}|^2+l_{23} |u|^2u^{(3)}_{xx}+l_{24} u^2 \bar u^{(3)}_{xx}+ l_{25} \bar u u_x u_{xxx}+\\
l_{26} u \bar u_x u_{xxx}+l_{27} u u_x \bar u_{xxx}+l_{28} |u|^2u_{xxxx}+ l_{29} u^2\bar u_{xxxx}\Big), 
\ea
\eeq
where
\beq
\ba{l}
l_{1}=-\frac{1}{18}c^3,~~l_{2}=-\frac{3}{2}c^2,~~l_{3}=2c,~~l_{4}=-c^2,~~l_{5}=4 c,~~l_{6}=7 l_{10},~~l_{7}=\frac{1}{2}l_{11}, \\ 
l_{8}=6 l_{10},~~l_{9}=-\frac{1}{3}(a_1  + a_2 + a_3 - 5 a_4 + 7 a_5 + a_6 + a_7 -11 a_8 + 13 a_9+ a_{10}), \\
l_{10}=-\frac{1}{36}c^2,~~l_{11}=-(a_1  + a_2 + a_3 + a_4 + a_5 - 3 a_6 + 5 a_7 -3 (a_8 + a_9)+ 5 a_{10}), \\ 
l_{12}=l_{13},~~l_{13}=-\frac{1}{3}(a_1  + a_2 + a_3 - 5 a_4 + 7 a_5 + a_6 + a_7 -11 a_8 + 13 a_9+ a_{10}), \\
l_{14}=5 l_{10},~~l_{15}=l_{16},      \\ 
l_{16}=-\frac{1}{3}(2 a_1  - a_2 + 2 a_3 - a_4 - a_5 - 4 (a_6 +a_7 +a_8 +a_9+a_{10}), \\
l_{17}=-\frac{1}{36}(5 (a_1  + a_2 + a_3 + a_4 + a_5 +a_6 + a_7+a_{10}) - 67 a_8 + 77 a_9), \\
l_{18}=-\frac{1}{18}(a_1  + a_2 + a_3 + a_4 + a_5 -8 (a_6 + a_7 + a_8 + a_9+a_{10})), \\
l_{19}=2 l_{10},~~l_{20}=-\frac{1}{3}(a_1  + a_2  + a_6 + a_7 - 5 (a_3+a_4+a_5+a_8 + a_9+a_{10})), \\
l_{21}=-\frac{1}{18}(a_1  + a_2 + a_3 + a_4 + a_5 + a_6 + a_7 -17 (a8 + a9)+ 19 a_{10}), \\ 
l_{22}=-\frac{1}{180}(11 (a_1+a_2 +a_3) - 79 (a_4+a_5) + 11 (a_6 +a_7) -169 (a_8 + a_9+a_{10})), \\
l_{23}=-\frac{1}{3}(2 a_1  - a_2 + 2 a_3 - a_4 - a_5 - 4(a_6+a_7+a_8+a_9+a_{10})),~~l_{24}=\frac{1}{2}l_{20}, \\
l_{25}=-\frac{1}{12}(a_1  + a_2 + a_3 + a_4 + a_5 + 9 a_7 - 7 (a_6+a_8 + a_9)+ 9 a_{10}), \\
l_{26}=-\frac{1}{60}(3(a_1  + a_2 +a_3+a_{10}) - 17 a_4 + 23 a_5 + 3 a_6 + 3 a_7 - 37 a_8 + 43 a_9), \\
l_{27}=-\frac{1}{45}(a_1  + a_2 + a_3 - 14 a_4 + 16 a_5 + a_6 + a_7 -29 a_8 + 31 a_9+ a_{10}), \\
l_{28}=-\frac{1}{60}(2 a_1- 3 a_2 + 2 a_3 - 3 a_4 - 3 a_5 - 8 (a_6 + a_7 + a_8 + a_9+a_{10})),  \\
l_{29}=-\frac{1}{180}(a_1 + a_2  + a_6 + a_7- 14 (a_3+a_4+a_5+a_8 + a_9+a_{10} )).
\ea
\eeq
The solution $f^{(3)}_4\in{\cal P}_{7}(1)$ of $\hat M_4 f^{(3)}_2=\hat M_2 f^{(3)}_4$, where $f^{(3)}_2=g_5$ is given in 
(\ref{g5}),(\ref{cj}), exists unique and reads
\beq\label{f43}
\ba{l}
f^{(3)}_4 = i\Big(\alpha_1 u |u|^6 + \alpha_2 u_{xx} |u|^4+ \alpha_3 \bar u_{xx} u^2|u|^2+ \alpha_4 {u^2_x} |u|^2\bar u 
+\alpha_5 |u_x|^2 |u|^2 u + \\
\alpha_6 \bar {u}^2_x u^3+\alpha_7 u_{xxxx} |u|^2+\alpha_8 \bar u_{xxxx}u^2+ \alpha_9 u_{xxx}u_x\bar u+ 
\alpha_{10} \bar u_{xxx}u_x u+ \alpha_{11} u_{xxx}\bar u_x u +\\
\alpha_{12} {u^2_{xx}}\bar u+  
\alpha_{13} |u_{xx}|^2u+\alpha_{14} u_{xx}|u_x|^2+ \alpha_{15}\bar u_{xx}{u_x}^2\Big),    
\ea
\eeq
where
\beq
\ba{l}
\alpha_1=\frac{c^2}{3} (2 c_2 - c_3 + c_4 + 3 c_5),~~\alpha_2=\frac{c}{6} (4 c_2 - 2 c_3 + 6 c_4 + 5 c_5), \\
\alpha_3=\frac{c}{12} (2 c_2 - c_3 + 3 c_4 + 10 c_5),~~\alpha_4=\frac{c}{24} (40 c_2 - 11 c_3 + 25 c_4 + 20 c_5), \\
\alpha_5=\frac{c}{12} (8 c_2 + 5 c_3 + 7 c_4 + 16 c_5),~~\alpha_6=\frac{c}{24} (4 c_2 + c_3 + c_4 + 8 c_5), \\
\alpha_7=\frac{c_4}{6},~~\alpha_8=\frac{c_5}{12},~~\alpha_9=\frac{1}{12} (4 c_2 + 3 c_4),~~
\alpha_{10}=\frac{1}{12} (c_3 + 2 c_5), \\
\alpha_{11}=\frac{1}{12} (2 c_3 + c_4),~~\alpha_{12}=\frac{1}{12} (3 c_2 + 2 c_4),~~\alpha_{13}=\frac{1}{12} (c_3 + c_4 + 4 c_5), \\
\alpha_{14}=\frac{1}{12} (2 c_2 + 5 c_3 + c_4),~~\alpha_{15}=\frac{1}{12}(c_2 + c_3 + 3 c_5),
\ea
\eeq
iff the following constraint is satisfied 
\beq\label{c1}
2 c_1-c(2 c_2 - c_3 + c4 + 4 c_5)=0
\eeq
on the coefficients $c_j$'s defined in (\ref{cj}). This constraint is equivalent to (\ref{constr1}). 

$f^{(5)}_2=g_7-f^{(3)}_4$ consequently reads, from $(\ref{g7})$ and $(\ref{f43})$:  
\beq\label{f25}
\ba{l}
f^{(5)}_2 = i\Big(d_1 u |u|^6+d_2 |u|^4 u^{(3)}+d_3 \bar u {u^{(3)}}^2+
d_4 u^2 |u|^2\bar u^{(3)}+d_5 u |u^{(3)}|^2+ \\
d_6 \bar u |u|^2{u^2_x}+d_7 {u^2_x} \bar u^{(3)}+d_8 u |u|^2 |u_x|^2+d_9 |u_x|^2 u^{(3)}+d_{10} u^3 \bar{u}^2_x +
d_{11} \bar u u_x u^{(3)}_x+ \\
d_{12}u \bar u_x u^{(3)}_x+d_{13} u u_x \bar u^{(3)}_x+ d_{14} |u|^4 u_{xx}+ d_{15} \bar u u_{xx}u^{(3)}+d_{16} u u_{xx}\bar u^{(3)}+\\
d_{17}|u_x|^2 u_{xx}+ d_{18}\bar u {u^2_{xx}}+d_{19}{u^2} |u|^2\bar u_{xx}+d_{20}u \bar u_{xx}u^{(3)}+ d_{21}{u^2_x} \bar u_{xx}+ 
d_{22} u |u_{xx}|^2 + \\
d_{23}|u|^2 u^{(3)}_{xx}+d_{24}u^2 \bar u^{(3)}_{xx}+d_{25}\bar u u_x u_{xxx}+ d_{26}u\bar u_x u_{xxx}+ d_{27}u u_x\bar u_{xxx}+ \\
d_{28}|u|^2u_{xxxx}+ d_{29}u^2\bar u_{xxxx}\Big), 
\ea
\eeq
where
\beq\label{dj}
\ba{l}
d_1=\frac{c^2}{9}(5 a_1  + 2 a_2 - 4 a_3 - a_4 - 13 a_5 - 13 a_6 + 11 a_7 - 10 a_8 - 34 a_9+ 2 a_{10}) , \\
d_2=-\frac{3}{2}c^2,~~d_3=2 c,~~d_4=-c^2,~~d_5=4 c, \\
d_6=\frac{c}{72}(95 a_1 + 20 a_2 + 35 a_3 + 26 a_4 - 106 a_5 - 295 a_6 + 185 a_7 -223 a_8 - 487 a_9+ \\
125 a_{10}),~~d_7=-\frac{1}{2}(a_1 + a_2 + a_3 + a_4 + a_5 - 3 a_6 + 5 a_7 -3 (a_8 + a_9)+ 5 a_{10}), \\
d_8=\frac{c}{12}(11 a_1  + 4 a_2 - 5 a_3 - 22 a_4 - 2 a_5 - 19 a_6 + 13 a_7 - 55 a_8 - 15 a_9- 3 a_{10}) ,  \\
d_9=d_{13},~~d_{10}=\frac{c}{72}(11 a_1 + 8 a_2 - 13 a_3 - 22 a_4 - 10 a_5 - 19 a_6 + 29 a_7 - 55 a_8 - \\
31 a_9 + 5 a_{10}) ,~~d_{11}= 2 d_7,~~d_{12}=d_{13},  \\
d_{13}=-\frac{1}{3}(a_1 + a_2 + a_3 - 5 a_4 + 7 a_5 + a_6 + a_7 -11 a_8 + 13 a_9+ a_{10}), \\
d_{14}=\frac{c}{18}(16 a_1 - 2 a_2 + a_3 - 5 a_4 - 29 a_5 - 44 a_6 + 4 a_7 - 35 a_8 -83 a_9 - 11 a_{10}), \\
d_{15}=d_{16},~~d_{16}=-\frac{1}{3}(2 a_1 - a_2 + 2 a_3 - a_4 - a_5 - 4(a_6+a_7+a_8+a_9+a_{10})),  \\
d_{17}=\frac{1}{36}(5 a_1 + 2 a_2 + 5 a_3 - 28 a_4 + 32 a_5 -13 a_6 + 11 a_7 - a_8 - 25 a_9+ 11 a_{10}), \\
d_{18}=\frac{1}{72}(13 a_1 + a_2 + 13 a_3 + a_4 + a_5 - 11 a_6 + 61 a_7 - 11 (a_8 + a_9)+ 61 a_{10}),  \\
d_{19}=\frac{c}{36}(11 a_1 + 2 a_2 - 19 a_3 - 22 a_4 - 34 a_5 - 19 a_6 + 5 a_7 -37 a_8 - 61 a_9 - 25 a_{10}), \\
d_{20}=-\frac{1}{3}(a_1 + a_2 + a_6 + a_7 - 5 (a_3+a_4+a_5+a_8 + a_9+a_{10})),  \\
d_{21}=\frac{1}{36}(2 a_1 + 2 a_2 - 7 a_3 - 13 a_4 - a_5 - 4 a_6 + 8 a_7 + 11 a_8 + 35 a_9- 37 a_{10}), \\
d_{22}=\frac{1}{180}(14 a_1 - a_2 - 46 a_3 - a_4 + 59 a_5 - 16 a_6 - 16 a_7 + 44 a_8 + 164 a_9+ \\
104 a_{10}), \\
d_{23}=-\frac{1}{3}(2 a_1 - a_2 + 2 a_3 - a_4 - a_5 - 4 (a_6+a_7+a_8 +a_9+a_{10})),~~d_{24}=\frac{1}{2}d_{20}, \\
d_{25}=\frac{1}{4}(a_1 + a_3 - a_6 - a_7 - a_8 - a_9- a_{10}), \\
d_{26}=\frac{1}{180}(11 a_1 - 4 a_2 + 11 a_3 - 4 a_4 - 4 a_5 -19 (a_6 + a_7 + a_8 + a_9+a_{10})), \\
d_{27}=\frac{1}{30}(a_1  + a_2 - 4 a_3 + a_4 - 9 a_5 + a_6 + a_7 + 6 a_8 - 14 a_9- 4 a_{10}), \\
d_{28}=\frac{1}{180}(14 a_1  - a_2 + 14 a_3 - a_4 - a_5 - 16 (a_6 + a_7 + a_8 + a_9+a_{10})), \\
d_{29}=\frac{c}{120}.
\ea
\eeq
The unique solution $f^{(3)}_6=\hat M_6 u^{(3)}\in{\cal P}_{9}(1)$ of equation $\hat M_6 f^{(3)}_2=\hat M_2 f^{(3)}_6$ 
reads:
\beq\label{f63}
\ba{l}
f^{(3)}_6=i\Big( \beta_1 |u|^8 u +\beta_2 u_{xx}|u|^6+\beta_3 \bar u_{xx}|u|^4u^2+\beta_4 {u^2_x}|u|^4\bar u+ \beta_5 |u_x|^2|u|^4u+\\ 
\beta_6 \bar {u}^2_x{u^3}|u|^2+ \beta_7 u_{xxxx}|u|^4+\beta_8 u_{xxx}u_x|u|^2\bar u+ \beta_9 u_{xxx}\bar u_x|u|^2u+ 
\beta_{10}{u^2_{xx}}|u|^2\bar u+ \\ 
\beta_{11} |u_{xx}|^2|u|^2u+\beta_{12}\bar u_{xxx}u_x|u|^2u+\beta_{13}\bar u_{xxx}\bar u_x{u^3}+
\beta_{14}u_{xx}{u^2_x}{\bar u}^2+ \\
\beta_{15}u_{xx}|u_x|^2|u|^2+\beta_{16}u_{xx}{\bar u}^2_x u^2+\beta_{17}{u^3_x}\bar u_x\bar u+\beta_{18}|u_x|^4u+ 
\beta_{19}(\bar u_{xx})^2u^3+ \\ 
\beta_{20}\bar u_{xx}|u_x|^2u^2+\beta_{21}\bar u_{xx}{u^2_x}|u|^2+ \beta_{22}u_{xxxxxx}|u|^2+ \beta_{23}\bar u_{xxxxxx}{u}^2+ \\
\beta_{24}u_{xxxxx}|u|^2+ 
\beta_{25}\bar u_{xxxxx}u u_x+\beta_{26}u_{xxxx}|u_x|^2+\beta_{27}u_{xxxx}u_{xx}\bar u + \\
\beta_{28}u_{xxxx}\bar u_{xx}u+ \beta_{29}\bar u_{xxxx}u^2_x+ 
\beta_{30}\bar u_{xxxx}u_{xx}u+\beta_{31}u^2_{xxx}\bar u+ \beta_{32}u_{xxx}u_{xx}\bar u_{x}+ \\
\beta_{33}u_{xxx}\bar u_{xx}u_{x}+ \beta_{34}|u_{xxx}|^2 u+ \beta_{35}|u_{xx}|^2 u_{xx}+ \beta_{36}u_{xxxxx}u_{x}\bar u+ \\
\beta_{37}u_{xx}\bar u_{xxx}u_{x}+\beta_{38}\bar u_{xxxx}|u|^2{u^2} \Big),
\ea
\eeq
where
\beq\label{betaj}
\ba{l}
\beta_1=\frac{c^3}{48}(6 c_2 - 3 c_3 + 3 c_4 + 8 c_5),~~\beta_2=\frac{c^2}{36}(10 c_2 - 5 c_3 + 10 c_4 + 12 c_5),~~ \\
\beta_3=\frac{c^2}{36}(4 c_2 - 2 c_3 + 4 c_4 + 9 c_5),~~\beta_{4}=\frac{c^2}{18}(13 c_2 - 5 c_3 + 9 c_4 + 10 c_5),~~ \\
\beta_5=\frac{c^2}{36}(18 c_2 - 3 c_3 + 14 c_4 + 28 c_5),~~\beta_6=\frac{c^2}{36}(5 c_2 - c_3 + 3 c_4 + 8 c_5),~~\\
\beta_7=\frac{c}{180}(6 c_2 - 3 c_3 + 15 c_4 + 7 c_5),~~\beta_8=\frac{c}{360}(108 c_2 - 29 c_3 + 126 c_4 + 56 c_5), \\
\beta_9=\frac{c}{360}(36 c_2 + 7 c_3 + 66 c_4 + 52 c_5),~~\beta_{10}=\frac{c}{720} (152 c_2 - 41 c_3 + 169 c_4 + 84 c_5), \\
\beta_{11}=\frac{c}{360}(44 c_2 - 7 c_3 + 79 c_4 + 118 c_5),~~\beta_{12}=\frac{c}{360}(16 c_2 + 7 c_3 + 26 c_4 + 62 c_5), \\
\beta_{13}=\frac{c}{360}(8 c_2 + c_3 + 6 c_4 + 26 c_5),~~\beta_{14}=\frac{c}{72}(33 c_2 - 7 c_3 + 21 c_4 + 14 c_5), \\
\beta_{15}=\frac{c}{360}(278 c_2 + 11 c_3 + 276 c_4 + 236 c_5),~~\beta_{16}=\frac{c}{360}(39 c_2 + 8 c_3 + 33 c_4 + 68 c_5), \\ 
\beta_{17}=\frac{c}{72}(22 c_2 - c_3 + 13 c_4 + 16 c_5),~~\beta_{18}=\frac{c}{720}(158 c_2 + 31 c_3 + 101 c_4 + 176 c_5), \\
\beta_{19}=\frac{c}{720}(12 c_2 - c_3 + 9 c_4 + 44 c_5),~~\beta_{20}=\frac{c}{120}(16 c_2 + 7 c_3 + 12 c_4 + 42 c_5), \\
\beta_{21}=\frac{c}{360}(114 c_2 - 12 c_3 + 103 c_4 + 158 c_5),~~\beta_{22}=\frac{c_4}{120},~~\beta_{23}=\frac{c_5}{360}, \\
\beta_{24}=\frac{1}{240}(2 c_3 + 3 c_4),~~\beta_{25}=\frac{1}{360}(c_3 + 4 c_5),~~
\beta_{26}=\frac{1}{720}(18 c_2 + 21 c_3 + 25 c_4),\\
\beta_{27}=\frac{1}{360}(15 c_2 + 13 c_4),~~\beta_{28}=\frac{1}{720}(9 c_3 + 11 c_4 + 12 c_5),~~\\
\beta_{29}=\frac{1}{360}(c_2 + 2 c_3 + 10 c_5),~~
\beta_{30}=\frac{1}{360}(2 c_3 + c_4 + 11 c_5),~~\\
\beta_{31}=\frac{1}{144}(4 c_2 + 3 c_4),~~\beta_{32}=\frac{1}{720}(50 c_2 + 35 c_3 + 34 c_4), \\
\beta_{33}=\frac{1}{360}(11 c_2 + 17 c_3 + 10 c_4 + 15 c_5),~~\beta_{34}=\frac{1}{720}(11 c_3 + 4 c_4 + 18 c_5),\\
\beta_{35}=\frac{1}{720}(20 c_2 + 25 c_3 + 11 c_4 + 20 c_5), \\
\beta_{36}=\frac{1}{240}(4 c_2 + 5 c_4),~~\beta_{37}=\frac{1}{720}(8 c_2 + 31 c_3 + 4 c_4 + 50 c_5),\\
\beta_{38}=\frac{c}{360}(2 c_2 - c_3 + 5 c_4 + 14 c_5),
\ea
\eeq
and no additional constraint on the coefficients $c_j$'s appears. \\ 
The unique solution $f^{(5)}_4=\hat M_4 u^{(5)}\in{\cal P}_{9}(3)$ of equation $\hat M_4 f^{(5)}_2=\hat M_2 f^{(5)}_4$
reads
\beq\label{f45}
\ba{l}
f^{(5)}_4=i\Big(\delta_1 u |u|^8+\delta_2 u_{xx}|u|^6+\delta_3 \bar u_{xx}u^2|u|^4+
\delta_4 u^2_x \bar u |u|^4+\delta_5 |u_x |^2u |u|^4+ \\
\delta_6 {\bar u}^2_x|u|^2 u^3 + \delta_7 u_{xxxx}|u|^4+\delta_8 u_{xxx}u_x |u|^2\bar u+\delta_9 u_{xxx}{\bar u}_x |u|^2u+
\delta_{10}u^2_{xx}|u|^2\bar u + \\
\delta_{11}|u_{xx}|^2|u|^2 u+\delta_{12}{\bar u}_{xxx}u_x |u|^2u+\delta_{13}{\bar u}_{xxx}{\bar u}_{x}u^3+
\delta_{14}u_{xx}u^2_x{\bar u}^2+\delta_{15}u_{xx}|u_x |^2|u|^2+ \\
\delta_{16}u_{xx}{\bar u}^2_xu^2+\delta_{17}u^3_x{\bar u}_x{\bar u}+\delta_{18}|u_x |^4u+\delta_{19}{\bar u}^2_{xx}u^3+
\delta_{20}{\bar u}_{xx}|u_x |^2u^2+\delta_{21}{\bar u}_{xx}u^2_x|u|^2+ \\
\delta_{22}u_{xxxxxx}|u|^2+\delta_{23}{\bar u}_{xxxxxx}u^2+\delta_{24}u_{xxxxx}{\bar u}_x u+
\delta_{25}{\bar u}_{xxxxx}{u}_x u+\delta_{26}u_{xxxx}|u_x |^2+ \\
\delta_{27}u_{xxxx}u_{xx}\bar u+\delta_{28}u_{xxxx}{\bar u}_{xx}u+
\delta_{29}{\bar u}_{xxxx}u^2_x+\delta_{30}{\bar u}_{xxxx}u_{xx}u+\delta_{31}u^2_{xxx}\bar u+ \\
\delta_{32}u_{xxx}u_{xx}{\bar u}_{x}+\delta_{33}u_{xxx}{\bar u}_{xx}u_x+\delta_{34}|u_{xxx}|^2u+
\delta_{35}u_{xx}|u_{xx}|^2+\delta_{36}u_{xxxxx}u_x\bar u+ \\
\delta_{37}u_{xx}{\bar u}_{xxx}u_x+\delta_{38}u^2|u|^2{\bar u}_{xxxx}+\gamma_1 u^{(3)}_{xxxx}|u|^2+\gamma_2 u^{(3)}_{xxx}u_x\bar u+ 
\gamma_3 u^{(3)}_{xxx}{\bar u}_{x}u+ \\
\gamma_4 u^{(3)}_{xx}u_{xx}\bar u+\gamma_5 u^{(3)}_{xx}|u_x |^2+\gamma_6 u^{(3)}_{xx}u {\bar u}_{xx}+
\gamma_7 u^{(3)}_{x}u_{xxx}\bar u+\gamma_8 u^{(3)}_{x}u_{xx}{\bar u}_{x}+ \\
\gamma_9 u^{(3)}_{x}u_{x}{\bar u}_{xx}+\gamma_{10} u^{(3)}_{x}u{\bar u}_{xxx}+\gamma_{11} u^{(3)}u_{xxxx}\bar u+
\gamma_{12} u^{(3)}u_{xxx}{\bar u}_{x}+\gamma_{13} u^{(3)}|u_{xx}|^2+ \\
\gamma_{14} u^{(3)}u_x{\bar u}_{xxx}+\gamma_{15} u^{(3)}u{\bar u}_{xxxx}+\gamma_{16}{\bar u^{(3)}}_{xxxx}u^2+
\gamma_{17}{\bar u^{(3)}}_{xxx}u_x u+\gamma_{18}{\bar u^{(3)}}_{xx}u_{xx} u+ \\
\gamma_{19}{\bar u^{(3)}}_{xx}u^2_x+\gamma_{20}{\bar u^{(3)}}_{x}u_{xxx}u+
\gamma_{21}{\bar u^{(3)}}_{x}u_{xx}u_{x}+\gamma_{22}{\bar u^{(3)}}u_{xxxx}u+\gamma_{23}{\bar u^{(3)}}u_{xxx}u_x+ \\
\gamma_{24}{\bar u^{(3)}}u^2_{xx}+\gamma_{25}u^{(3)}_{xx}|u|^4+\gamma_{26}u^{(3)}_{x}u_x|u|^2\bar u+
\gamma_{27}u^{(3)}_{x}{\bar u}_{x}|u|^2u+ \\
\gamma_{28}{u^{(3)}}u_{xx}|u|^2\bar u+ \gamma_{29}{u^{(3)}}u^2_x{\bar u}^2+\gamma_{30}u^{(3)}|u_x|^2|u|^2+
\gamma_{31}u^{(3)}{{\bar u}_{x}}^2u^2+ \gamma_{32}u^{(3)}{{\bar u}_{xx}}|u|^2u+ \\
\gamma_{33}{\bar u^{(3)}}_{xx}|u|^2u^2+\gamma_{34}{\bar u^{(3)}}_{x}u_x|u|^2u+
\gamma_{35}{\bar u^{(3)}}_{x}{{\bar u}_{x}}u^3+\gamma_{36}{\bar u^{(3)}}u_{xx}|u|^2u+
\gamma_{37}{\bar u^{(3)}}u^2_x|u|^2+ \\
\gamma_{38}{\bar u^{(3)}}|u_x |^2u^2+\gamma_{39}{\bar u^{(3)}}{{\bar u}_{xx}}u^3+
\gamma_{40}u^{(3)}|u|^6+\gamma_{41}{\bar u^{(3)}}|u|^4u^2+\sigma_1 u^{(3)}_{xx}{u^{(3)}}{\bar u}+ \\
\sigma_2 {u^{(3)}_{x}}^2\bar u+\sigma_3 u^{(3)}_{x}{u^{(3)}}{\bar u}_x+\sigma_4 {u^{(3)}}^2{\bar u}_{xx}+ 
\sigma_5 u^{(3)}_{xx}{\bar u}^{(3)}u+\sigma_6 |u^{(3)}_{x}|^2u+ \\
\sigma_7 u^{(3)}_{x}{\bar u}^{(3)}u_x+
\sigma_8 {u^{(3)}}{\bar u^{(3)}}_{xx}u+\sigma_9 {u^{(3)}}{\bar u^{(3)}}_{x}u_x+\sigma_{10}|{\bar u^{(3)}}|^2u_{xx}+ 
\sigma_{11}({u}^{(3)})^2 |u|^2\bar u+ \\
\sigma_{12}|{\bar u^{(3)}}|^2|u|^2 u+\sigma_{13}(\bar u^{(3)})^2 u^3 \Big),
\ea
\eeq
where:
\beq\label{coeff_f45a}
\ba{l}
\delta_1=\frac{c}{576} \Big(336 d_1+ 4(6 c_2- 3 c_3+ 2 c_4+ 10 c_5) d_2+ 
6c_5(2 c_2- c_3+ 2 c_4+ 4 c_5) d_3+ \\
4(6 c_2- 3 c_3+ 3 c_4+ 2 c_5) d_4+ (- 4 c^2_4+ 2 c_2 c_5- c_3 c_5+ c_4 c_5- 2 c^2_5) d_5+ \\ 
c( 2(14 c_2- 5 c_3+ 9 c_4+ 28 c_5) d_{11}-2(10 c_2- 3 c_3+7 c_4+8 c_5) d_{12}-4( 6 c_2- \\
3 c_3+c_4 +16 c_5) d_{13} + 16 d_{14}+ 2(18 c_2- 7 c_3+ 11 c_4+ 30 c_5) d_{15} -2( 6 c_2- \\
3 c_3+3 c_4 +14 c_5) d_{16} - 48 d_{19}-4(6 c_2-3 c_3+3 c_4+8 c_5) d_{20}+4(- 4 c_2+  c_3+ \\
2 c_4+ 15 c_5) d_{23}+4(- 6 c_2+ 5 c_3+ c_4- 6 c_5) d_{24} + 32 d_6- 16 d_8+ 4(2 c_2- c_3+ \\
c_4+4 c_5) d_9) +8c^2(- 2 d_{17}+ 4 d_{18}+ 2 d_{22}+ 7 d_{25}- 7 d_{26}- 5 d_{28}+ 16 d_{29})\Big), \\
\delta_{2}=\frac{1}{144}\Big(96 d_1 + 10 c_4 d_2 + 4 c_4 c_5 d_3 - 6 c_5 d_4 - 2 c^2_4 d_5 +c(2(4 c_2- c_3+ 6 c_4+ \\
8 c_5) d_{11}-2(2 c_2+ 5 c_4) d_{12}-2(6 c_2- 3 c_3 + c_4 + 14 c_5) d_{13}+ 104 d_{14}+ (18 c_2 - \\
7 c_3 + 11 c_4+ 24 c_5) d_{15}+3(- 2 c_2 + c_3-c_4- 4 c_5) d_{16}- 24 d_{19}+ (34 c_2- 19 c_3+ \\
31 c_4+ 80 c_5) d_{23}+2(-6 c_2 + 5 c_3- 5 c_4- 10 c_5) d_{24}+ 16 d_6- 8 d_8+ 2(2 c_2- \\ 
c_3+c_4+ 4 c_5) d_9)+4 c^2(- 2 d_{17}+ 4 d_{18}+ 2 d_{22}+ 13 d_{25}- 13 d_{26}+ d_{28}+ 4 d_{29})\Big),  \\
\delta_{3}=\frac{1}{48} (12 d_1 + 2 c_5 d_2 + 2 c_4 d_4 + c(2 c_5 d_{11} -2 c_5 d_{12}+ 8 d_{14}+ 24 d_{19}+(2 c_2- \\
c_3+ c_4+4 c_5) d_{20}+ 2(2 c_2- c_3 + c_4 + 5 c_5) d_{23}+ 5(2 c_2- c_3 + c_4) d_{24})+ \\
8 c^2( d_{25}-d_{26}+ d_{28}- 2 d_{29})), \\
\delta_{4}=\frac{1}{144}\Big(216 d_1 + 2(3 c_2+ 6 c_4- 2 c_5) d_2 + 2(2 c_3 c_4+ 2 c_2 c_5+ c_3 c_5+ c_4 c_5- \\
2 c^2_5) d_3 - 24 c_5 d_4 - (6 c^2_4- 3 c_3 c_5+ c_4 c_5+ 2 c^2_5) d_5 +
c(-24 d_{10}+(52 c_2- 19 c_3+ \\
9 c_4+ 54 c_5) d_{11}+2(- 7 c_2+ 3 c_3- 8 c_4+ 9 c_5) d_{12} +2(- 10 c_2+ 4 c_3- 5 c_4- \\
22 c_5) d_{13}+ 28 d_{14}+ 2(33 c_2- 4 c_3+ 5 c_4+ 16 c_5) d_{15} +2(- 22 c_2+ 12 c_3- 18 c_4- \\
19 c_5) d_{16}- 102 d_{19}+ (4 c_2- 10 c_3- 6 c_4+ 7 c_5) d_{20} + (154 c_2- 58 c_3+ 121 c_4 + \\
199 c_5) d_{23}+3(- 22 c_2+ 18 c_3- 3 c_4- 22 c_5) d_{24} + 132 d_6+ (10 c_2- 5 c_3+ 5 c_4+ \\
4 c_5) d_7- 16 d_8+ 2(6 c_2- 3 c_3+ 3 c_4+ 13 c_5) d_9)+2c^2(- 42 d_{17}+ 118 d_{18}- 8 d_{21}+ \\
17 d_{22}+ 54 d_{25}- 37 d_{26}+ 7 d_{27}+ 193 d_{28}+ 160 d_{29})\Big),                        \\
\delta_{5}=\frac{1}{48}\Big(48 d_1 + 2(3 c_3- 2 c_4+ 4 c_5) d_2 + 2 c_4 c_5 d_3 +2(- c_3-6 c_4+ 4 c_5) d_4 + \\ 
2(c^2_4 -c^2_5) d_5 +c(- 8 d_{10}+ 4(c_2+ 4 c_5) d_{11}+ (10 c_2- 7 c_3+ 11 c_4+ 8 c_5) d_{12}+   \\ 
(6 c_2- 3 c_3+ 3 c_4+ 4 c_5) d_{13}+24 d_{14}-4(3 c_2- c_3+c_4+ c_5) d_{15}+ 4(4 c_2- \\
2 c_3+ 3 c_4+ 3 c_5) d16+ 16 d_{19}+ 4 c_5 d_{20}+2(28 c_2- 15 c_3+ 6 c_4+ 42 c_5) d_{23}+ \\
4(- 4 c_2+ c_3+ 5 c_4- 17 c_5) d_{24}+ 8 d_6+ 32 d_8+(- 2 c_2+ c_3- c_4 - 4 c_5) d_9)+  \\
4c^2(2 d_{17} + 4 d_{18}- 6 d_{22}+ 3 d_{25}+ d_{26}- 4 d_{27}- 13 d_{28}- 4 d_{29})\Big),          \\
\delta_{6}=\frac{1}{48}\Big(12 d_1 + 2 c^2_5 d_3 + 2(c_2-c_4) d_4 - c_4 c_5 d5 + c(20 d_{10}+ 2(2 c_2- \\ 
c_3+ 4 c_5) d_{12}- 4 d_{14}+ 2(c_2+2 c_5) d_{15}- 2(c_4+c_5) d_{16}+ 8 d_{19}+ (8 c_2- \\
2 c_3+ 7 c_4+ 8 c_5) d_{23}+ 2(10 c_2- 5 c_3+ 11 c_4+ 5 c_5) d_{24}+ 4 d_6+ 4 d_8) + \\
2 c^2(- 2 d_{17}+ 12 d_{18}- 2 d_{22}+ d_{25}+3 d_{26}- 4 d_{27}+ 9 d_{28}+ 12 d_{29}) \Big), \\
\delta_{7}=\frac{1}{24} \Big(4 d_{14} + 2 c_4 d_{23} - c_5 d_{24}+2c( d_{25}-d_{26}+ 9 d_{28}- 2 d_{29})\Big), \\
\delta_{8}=\frac{1}{24}\Big(c_4 d_{11} - c_5 d_{13} + 8 d_{14} + 2 c_2 d_{15} + (6 c_2 + 7 c_4) d_{23} - 2 c_5 d_{24} + \\
 8 d_6 +2c(- 2 d_{17}+ 8 d_{18}+ 12 d_{25}- 2 d_{26}- d_{27}+ 27 d_{28}- 4 d_{29})\Big), \\
\delta_{9}=\frac{1}{24}\Big(c_4 d_{12} + 4 d_{14} + 2 c_5 d_{15} - 2 c_4 d_{16} + (3 c_3+ 4 c_4)d_{23} - c_3 d_{24} + 4 d_8+\\
2c(4 d_{18}- 2 d_{22}+ d_{25}+ 9 d_{26}-d_{27}+ 18 d_{28})\Big), \\
\delta_{10}=\frac{1}{24}\Big(2 c_4 d_{11} - 2 c_5 d_{13} + 8 d_{14} + 3 c_4 d_{15} - c_5 d_{16} + 2 c_2 d_{23} + 
c_4 d_{23} + 6 d_6 + \\
2c( 11 d_{18}-d_{22}+ 2 d_{25}- 2 d_{27}+ 6 d_{28}) \Big), 
\ea
\eeq
\beq\label{coeff_f45b}
\ba{l}
\delta_{11}=\frac{1}{24} \Big(2 c_5 d_{11} - 2 c_4 d_{13} + 4 d_{14} + c_5 d_{15} + c_4 d_{16} + 12 d_{19} + c_4 d_{20} 
+ (4 c_3+ c_4+ \\
6 c_5) d_{23} +(- c_4+ 2 c_5) d_{24}+ 2 d_8+4c( d_{18}+ 4 d_{22}+ d_{25}- d_{26}+ 13 d_{28}- 2 d_{29})\Big),     \\
\delta_{12}=\frac{1}{24}\Big(c_5 d_{11}+ c_4 d_{13} + 6 d_{19} + c_3 d_{23} + 10 c_5 d_{23} + c_3 d_{24} - 4 c_4 d_{24} + 
 2 d_8 + \\2c( d_{25} + d_{26}+ 7 d_{27}+ 8 d_{28} + 8 d_{29})\Big),    \\
\delta_{13}=\frac{1}{24} \Big(4 d_{10} + c_5 d_{12} + 2 d_{19} + 2 c_2 d_{24} + 2 c_4 d_{24}+
2c(d_{25} + d_{26}+d_{27}+ 8 d_{29})\Big), \\
\delta_{14}=\frac{1}{48}\Big(4(3 c_2+c_4) d_{11} + 8 d_{14} + 2(2 c_2 - c_4) d_{15} - 2 c_5 d_{16} + (12 c_2+ 7 c_4) d_{23} +\\ 
2 c_5 d_{24} + 24 d_6 - 6 c_5 d_7 +2c(- 4 d_{17}+ 20 d_{18}- 6 d_{21}-2 d_{22}+ 37 d_{25}- 2 d_{26}+ \\
37 d_{28}+ 4 d_{29})\Big),\\
\delta_{15}=\frac{1}{24}\Big((5 c_3+ c_4) d_{11} + (2 c_2+ c_4) d_{12} +(- 3 c_3+ 2 c_5) d_{13} +8 d_{14} + (c_3+ \\
4 c_5) d_{15} + (c_3- 4 c_4) d_{16} + (8 c_2+ 11 c_3+ 6 c_4- 4 c_5) d_{23} +4(- c_3+ c_4- \\
2 c_5) d_{24} + 8 d_6 - 4 c_4 d_7 + 20 d_8 + c_4 d_9+4c(4 d_{17}+ 8 d_{18}+ 11 d_{25}+ 5 d_{26}- 7 d_{27}+ \\
34 d_{28}-8 d_{29})\Big), \\
\delta_{16}=\frac{1}{48}\Big(12 d_{10} - 2 c_5 d_{11} + 2(2 c_3 + c_4 + 3 c_5) d_{12} - 2(2 c_2+ c_4) d_{13}+ \\
4 d_{14}+ 2 c_2 d_{16} + 
6 d_{19} - 3 c_5 d_{20} - (2 c_3 +7 c_5) d_{23} +(4 c_2 - 2 c_3 + c_4 + \\
4 c_5) d_{24} +8 d_8 + 2 c_5 d_9+ 2c(2 d_{17}+ 6 d_{18}+ 2 d_{21}+ d_{22}- 3 d_{25}+ 26 d_{26}-11 d_{27}- \\
14 d_{28}+ 4 d_{29})\Big), \\
\delta_{17}=\frac{1}{24}\Big((c_2+ 3 c_3) d_{11} - 2 c_3 d_{13} + 2( c_2+ c_3- c_5) d_{23} + 2 c_4 d_{24} + 4 d_6 + \\ 
(c_3- 2 c_4) d_7 + 6 d_8 + c_2 d_9 + 4c( d_{17}+ 2 d_{21}+ 7 d_{25}- 4 d_{27}+ 6 d_{28})\Big),  \\
\delta_{18}=\frac{1}{24}\Big(6 d_{10} - 2 c_5 d_{11} + (c_2 + 3 c_3 + 6 c_5) d_{12} - (3 c_2 - c_3 +2 c_4) d_{13} + \\
6 d_{19} - 3 c_5 d_{20} - (2 c_3 +7 c_5) d_{23} + (2 c_2 - 6 c_3 + c_4) d_{24} + 2 d_6 + c_2 d_7 + \\
8 d_8 + c_3 d_9 + 2 c_5 d_9 + 2c(4 d_{17}+ 6 d_{18}+ 2 d_{21}- 3 d_{22}- 3 d_{25}+20 d_{26}- 5 d_{27}\\
- 14 d_{28}- 20 d_{29})\Big), \\
\delta_{19}=\frac{1}{24} \Big(2 d_{10} + 2 d_{19} + c_5 d_{20} + (2 c_2 + c_4) d_{24}+
2c( d_{18}+d_{22}+ 6 d_{29})\Big),       \\
\delta_{20}=\frac{1}{24}\Big(24 d_{10} + (c_3 + 2 c_5) d_{12} + 2 c_2 d_{13} + c_4 d_{13} + 12 d_{19} 
+ (c_3+ \\
3 c_5) d_{20} - 3 c_5 d_{23} - (14 c_2 - 3 c_3+ 7 c_4) d_{24} + 4 d_8 + c_5 d_9 +2c(2 d_{17}- \\
6 d_{18}+ 2 d_{21}+ 7 d_{22}- 3 d_{25}+ 6 d_{26}+ 9 d_{27}- 36 d_{29})\Big), \\
\delta_{21}=\frac{1}{24}\Big((c_3+ 6 c_5)d_{11} + (c_3- 3 c_4)d_{13} + 18 d_{19} + c_2 d_{20} +(c_2+ 6 c_3+ 10 c_5)d_{23}-\\ 
4(c_3+c_4) d_{24} + 4 d_6 + c_4 d_7 + 4 d_8+4c(4 d_{21}+d_{22}+5 d_{25}- 2 d_{26}+ 
 3 d_{27}+ 20 d_{28}- \\
16 d_{29})\Big), \\
\delta_{22}=\frac{d_{28}}{6},~~\delta_{23}=\frac{d_{29}}{12},~~\delta_{24}=\frac{1}{12} (2 d_{26} + d_{28}), ~~ 
\delta_{25}=\frac{1}{12} (d_{27} + 2 d_{29}), \\
\delta_{26}=\frac{1}{12} (2 d_{17} + d_{25} + 3 d_{26} + d_{28}), ~~ 
\delta_{27}=\frac{1}{12} (4 d_{18} + 3 d_{25} + 2 d_{28}), \\ 
\delta_{28}=\frac{1}{12} (2 d_{22} + d_{26} + d_{28}), ~~
\delta_{29}=\frac{1}{12} (d_{21} + d_{27} + 3 d_{29}), \\
\delta_{30}=\frac{1}{12} (d_{22} + d_{27} + 4 d_{29}), ~~ 
\delta_{31}=\frac{1}{12} (3 d_{18} + 2 d_{25}), \\
\delta_{32}=\frac{1}{12}(5 d_{17} + 2 d_{18} + d_{25} + 2 d_{26}), ~~
\delta_{33}=\frac{1}{12} (d_{17} + 4 d_{21} + 3 d_{22} + d_{25} + d_{26}), \\
\delta_{34}=\frac{1}{12} (d_{22} + d_{26} + 2 d_{27}), ~~
\delta_{35}=\frac{1}{12} (d_{17} + d_{18} + 3 d_{21} + 2 d_{22}), \\ 
\delta_{36}=\frac{1}{12} (2 d_{25} + 3 d_{28}), ~~
\delta_{37}=\frac{1}{12} (d_{17} + 2 d_{21} + d_{22} + 5 d_{27}), \\
\delta_{38}=\frac{1}{24} (2 d_{19} + c_5 d_{23} + c_4 d_{24}+4c(d_{28}+3 d_{29})), \\
\ea
\eeq
\beq
\ba{l}
\gamma_{1}=\frac{d_{23}}{6}, ~~
\gamma_{2}=\frac{1}{12} (2 d_{11} + 3 d_{23}), ~~
\gamma_{3}=\frac{1}{12} (2 d_{12} + d_{23}), \\
\gamma_{4}=\frac{1}{12}(3 d_{11} + 2 d_{15} + 2 d_{23}), ~~
\gamma_{5}=\frac{1}{12}(d_{11} + 3 d_{12} + d_{23} + 2 d_9), \\
\gamma_{6}=\frac{1}{12} (d_{12} + 2 d_{20} + d_{23}), ~~
\gamma_{7}=\frac{1}{12}(2 d_{11} + 3 d_{15}), \\
\gamma_{8}=\frac{1}{12}(d_{11} + 2 d_{12} + d_{15} + 3 d_9), ~~
\gamma_{9}=\frac{1}{12} (d_{11} + d_{12} + 3 d_{20} + d_9), \\
\gamma_{10}=\frac{1}{12}(d_{12} + d_{20}), ~~
\gamma_{11}=\frac{d_{15}}{6}, ~~
\gamma_{12}=\frac{1}{12}(d_{15} + 2 d_9), \\
\gamma_{13}=\frac{1}{12} (d_{15} + 2 d_{20} + d_9), ~~
\gamma_{14}=\frac{1}{12}(d_{20} + d_{9}), ~~
\gamma_{15}=\frac{d_{20}}{12}, \\
\gamma_{16}=\frac{d_{24}}{12}, ~~
\gamma_{17}=\frac{1}{12} (d_{13} + 2 d_{24}), ~~
\gamma_{18}=\frac{1}{12} (d_{13} + d_{16} + 4 d_{24}), \\
\gamma_{19}=\frac{1}{12} (d_{13} + 3 d_{24} + d_7), ~~
\gamma_{20}=\frac{1}{12} (2 d_{13} + d_{16}), ~~
\gamma_{21}=\frac{1}{12}(5 d_{13} + d_{16} + 2 d_7), \\
\gamma_{22}=\frac{d_{16}}{6}, ~~
\gamma_{23}=\frac{1}{12}(3 d_{16} + 4 d_7), ~~
\gamma_{24}=\frac{1}{12}(2 d_{16} + 3 d_7), \\
\gamma_{25}=\frac{1}{12}(2 d_2 + c(d_{11}- d_{12}+ 9 d_{23}- 2 d_{24})), \\
\gamma_{26}=\frac{1}{12}\Big(6 d_2 + 2 c_4 d_3 - c_5 d_5 + c(11 d_{11}- d_{12}- d_{13}+ 2 d_{15}- 2 d_{20}+ 10 d_{23}- \\ 
8 d_{24})\Big), \\
\gamma_{27}=\frac{1}{12}(2 d_2 + 2 c_5 d_3 - c_4 d_5+c(d_{11}+ 7 d_{12}+ d_{13}+ 2 d_{15}-2 d_{16}+ 8 d_{23})), \\
\gamma_{28}=\frac{1}{24}\Big(8 d_2 + 6 c_4 d_3 - c_5 d_5 + 2c(2 d_{11}- 2 d_{13}+ 11 d_{15}- d_{16}- d{20}+ 2 d_{23})\Big),\\
\gamma_{29}=\frac{1}{24}\Big(6 d_2 + 2(c_2+ c_3-2 c_5) d_3 - (c_3-c_4) d_5 + 2c(4 d_{11}- 2 d_{13}+ 4 d_{15}- \\
2 d_{20}+ 2 d_{23}+  d_7+  d_9)\Big),                        \\
\gamma_{30}=\frac{1}{24}\Big( 8 d_2 + 2(c_3+ 4 c_5) d_3 + (c_3- 4 c_4) d_5 + 4c(d_{11}+ d_{12}+d_{13}+ 4 d_{15}-\\ 
2 d_{16}+ 2 d_{20}+ 6 d_{23}- 4 d_{24}+ 4 d_9)\Big),\\
\gamma_{31}=\frac{1}{24}(2 d_2 + c_2 d_5+2c(2 d_{12}+ 2 d_{20}+ 2 d_{24}+d_7+d_9)), \\
\gamma_{32}=\frac{1}{24}(4 d_2 + 2 c_5 d_3 + c_4 d_5+ 2c(d_{15}+ d_{16}+7 d_{20}+ 2 d_{23}+ 2 d_{24})), \\
\gamma_{33}=\frac{1}{2}(d_{4}+2c(d_{23}+3 d_{24})), \\
\gamma_{34}=\frac{1}{12} (3 d_4 + c(d_{11}+ d_{12}+ 7 d_{13}+ 4 d_{23}+ 4 d_{24})), \\
\gamma_{35}=\frac{1}{12}(d_{4}+c(d_{11}+ d_{12}+ d_{13}+ 4 d_{24})), \\
\gamma_{36}=\frac{1}{24}\Big(12 d_4 + c_4 d_5 + 2c(2 d_{11}- 2 d_{13}+ 3 d_{15}+ 7 d_{16}- d_{20}+ 6 d_{23}- 2 d_{24})\Big),\\
\gamma_{37}=\frac{1}{24}\Big(18 d_4 + c_2 d_5 + 4c(3 d_{11}- 2 d_{12}+d_{13}+d_{16}+ 7 d_{23}-8 d_{24}+ 4 d_7\Big), \\
\gamma_{38}=\frac{1}{24}\Big(18 d_4 + c_3 d_5 +4c(d_7 + d_9- 3 d_{11}+ 4 d_{12} + d_{13}+ 2 d_{16}- 3 d_{23} - 10 d_{24})\Big), \\
\gamma_{39}=\frac{1}{24}(2 d_4 + c_5 d_5+2c( d_{15}+d_{16}+d_{20}+ 2 d_{24})), \\
\gamma_{40}=\frac{c}{144} \Big(80 d_2 + 2(6 c_2- 3 c_3+  c_4+ 12 c_5) d_3 + 16 d_4 + (10 c_2- 5 c_3+ 5 c_4+ \\
8 c_5) d_5 + 8c(3 d_{11}-d_{12}- 2 d_{13}+ 3 d_{15}-d_{16}- 2 d_{20}+ 2 d_{23}-2 d_{24})\Big), \\
\gamma_{41}=\frac{c}{48} \Big(8 d_2 + 24 d_4 + (2 c_2 - c_3+ c_4+ 4 c_5) d_5 + 8c( d_{11}- d_{12}+ d_{23}- 2 d_{24})\Big), \\
\sigma_1=\frac{d_{3}}{3}, ~~
\sigma_2=\frac{d_{3}}{4},~~ \sigma_3=\frac{d_{3}}{6}, ~~
\sigma_4=\frac{d_{3}}{12}, ~~
\sigma_5=\sigma_{10}=\frac{d_{5}}{6}, ~~\sigma_6=\frac{d_{5}}{12},~~\sigma_7=\frac{d_{5}}{4}, \\ 
\sigma_8=\sigma_9=\frac{d_{5}}{12},~~
\sigma_{11}=\frac{c}{12}(7 d_3 + d_5), ~~ \sigma_{12}=\frac{c}{6} (d_3 + 4 d_5), ~~ 
\sigma_{13}=\frac{c}{12} (d_3 + d_5), 
\ea
\eeq
iff the following three sets of constraints are satisfied:
\beq\label{constr1_eps4}
\ba{l}
d_{15} - d_{16}=d_5 - 2 d_3=c_4 d_3 - 2 c d_{23}=-d_{12} + d_{13} - d_{15} + d_{20} + d_{23} - 2 d_{24}= \\
-d_9 + d_{11} + d_{13} - 2 d_7=(2 c_2 - c_3 + c_4) d_3 - 2 (d_{11} - d_{13} + d_{16} - d_{20} + 2 d_{24})c =  \\
(c_2 - c_3 + c_4 + c_5) d_3 + 2 (d_{13} - d_{16} - d_{24} - d_7)c=0,
\ea
\eeq
\beq\label{constr2_eps4}
(2 c_2 - c_3 + c_4 + 2 c_5) d_3 - 2 d_4 + 4 d_{24}c=-4 d_2 + (6 c_2 - 3 c_3 + 3 c_4 + 8 c_5) d_3 + 8 d_{24} c=0,
\eeq
\beq
\ba{l}\label{constr3_eps4}
-4 d_{10} + 2 c_5 d_{13} - 2 c_5 d_{16} + 2 d_{19} + c_5 d_{20} + c_5 d_{23} + (2 c_2- c_4- 4 c_5) d_{24} + \\ 
2c(2 d_{18}- d_{22}- d_{25}+ 2 d_{26}- d_{27}- 2 d_{28}+ 4 d_{29})=0, \\
\ \\
- 4 d_{14} -2 c_4 d_{11} + 2 (c_2 - c_4 + c_5) d_{16} + (4 c_2 + 3 c_4) d_{23} + 2 c_5 d_{24}+2c(- 2 d_{17} + \\ 
4 d_{18}+ 2 d_{22} + d_{25} - d_{26} + 13 d_{28} + 4 d_{29})=0, \\
\ \\
-2 d_8-(c_4 - 2 c_5) d_7  -2 (c_4 - c_5) d_{11} + (2 c_2 + c_3 + 2 c_4 + c_5) d_{13} -(c_3 + 3 c_4) d_{16} + \\
6 d_{19}+(- c_3+ 2 c_4+ c_5) d_{20} + (5 c_2 - c_3 + 3 c_4) d_{23} - (4 c_2 + c_3 + 6 c_4) d_{24}  + \\
2c (2 d_{17} - 2 d_{18} - 4 d_{21} - 7 d_{22} - d_{25} + 2 d_{26} + 11 d_{27} + 10 d_{28} - 20 d_{29})=0, \\
\ \\
- 4 d_6 + (2 c_2 + 2 c_3 - c_4 - 4 c_5) d_7 -12 d_{10} + (c_3 + 2 c_5) d_{11} - (c_2 - 4 c_4 + 9 c_5) d_{13}- \\ 
4 d_{14} + (3 c_2 - 5 c_5) d_{16} + 12 d_{19} -(2 c_2+c_3+2 c_4 - 4 c_5) d_{20} + (c_2 + 4 c_3 + c_4 - \\
c_5) d_{23} + (6 c_2 - 2 c_3 + 3 c_4 - 12 c_5) d_{24} +(- 8 d_{17} + 20 d_{18} + 4 d_{21} - 14 d_{22} + 10 d_{25} + \\
4 d_{26} + 10 d_{27} + 44 d_{28} - 16 d_{29})=0, \\
\ \\
-24 d_1 + 6 (c^2_5-c_4 c_5) d_3 + 2(6 c_2 - c_4) d_4 + ( 2(2 c_2 + 2 c_3 - c_4 + 4 c_5) d_7 - 4 d_8 - \\
52 d_{10}+2(- 4 c_2 + c_3 - c_4 + 2 c_5) d_{11} + 2 (c_2 +  c_3 - 6 c_4 + 6 c_5) d_{13} - 16 d_{14} + 2(5 c_2 - \\
c_3 + c_4 - c_5) d_{16} + 62 d_{19} +(- 4 c2 + 2 c_3- 12 c_4 + 5 c_5) d_{20} + (2 c_2 + 12 c_3 + 4 c_4 + \\
13 c_5) d_{23} + (6 c_2 - 12 c_3 + 5 c_4) d_{24})c +2 c^2(- 16 d_{17} + 50 d_{18} + 4 d_{21} - 13 d_{22} + d_{25} + \\
20 d_{26} + 3 d_{27} + 52 d_{28} + 4 d_{29})=0.
\ea
\eeq
The first set of constraints (\ref{constr1_eps4}) is automatically satisfied by the parametrizations (\ref{cj}) and (\ref{dj}),  
while the second set of two constraints (\ref{constr2_eps4}) is satisfied by the parametrizations (\ref{cj}) and (\ref{dj}) 
and by the $O(\eps^2)$ constraint (\ref{constr1}). The remaining   
five constraints (\ref{constr3_eps4}) are equivalent to the five quadratic 
constraints (\ref{S4}),(\ref{defQ1})-(\ref{defQ5}) in the S-integrability scenario in which (\ref{Sint_1}) holds, and to the 
linear constraints (\ref{Cint_3}) in the C-integrability scenario in which $c=0$ (the last constraint is automatically 
satisfied by the condition $c=0$ and, in the remaining constraints, the quadrics degenerate into the hyperplanes described 
by equations (\ref{Cint_3})).          

\end{footnotesize}

\vskip 10pt
\noindent
{\bf Acknowledgements}. We acknowledge interesting discussions with U. Aglietti.



\begin{thebibliography}{9}

\bibitem{DMS} A. Degasperis, S. V. Manakov and P. M. Santini, ``Multi-scale perturbation beyond the nonlinear 
Schr\"odinger equation. I'', Physica D {\bf 100}, 187-211 (1997).

\bibitem{DP} A. Degasperis and M. Procesi, ``Asymptotic integrability'' in {\it Simmetry and perturbation 
theory, SPT98} 23-37, edited by A. Degasperis and G. Gaeta, World Scientific, Singapore (1999).

\bibitem{Degasperis} A. Degasperis, ``Multiscale expansion and integrability of dispersive wave equations'', lectures given 
at the Euro Summer School ``What is integrability?'', Isaac Newton Institute, Cambridge, U.K., 
13-24 August (2001); in {\it Integrability} edited by A. Mikhailov, Lecture Notes in Physics {\bf 767}, Springer, 
Berlin-Heidelberg (2009). 

\bibitem{W} P. Winternitz, ``Symmetries of discrete systems.  Discrete integrable systems'', 185--243, {\it Lecture Notes 
in Phys. 644}, eds. B. Grammaticos, Y. Kossmann-Schwarzbach and T. Tamizhmani, Springer, Berlin, 2004.

\bibitem{VW} F. Valiquette and P. Winternitz, ``Discretization of partial differential equations preserving 
their physical symmetries'', arXiv:math-ph/0507061.

\bibitem{TA} T. R. Taha and M. J. Ablowitz, ``Analytical and numerical aspects of certain nonlinear evolution equations II. 
Numerical, nonlinear Schrodinger equation'', J. Comput. Phys. 55, 192 (1984). 

\bibitem{HA} B. M. Herbst and M. J. Ablowitz, ``Numerically induced chaos in the nonlinear Schr\"odinger equation'', 
Phys. Rev. Lett. {\bf 62}, 2065-2068 (1989).

\bibitem{Kelley} P. L. Kelley, Phys. Rev. Lett. (1965) {\bf 15} 1005.

\bibitem{Zakharov} V. E. Zakharov, Soviet Phys. JETP (1968) 994-998.

\bibitem{BN} D. J. Benney and A. C. Newell, J. Math. and Phys. (now Stud. Appl. Math.) {\bf 46} (1967) 133-139. 

\bibitem{HT} A. Hasegawa and T. Tappert, Appl. Phys. Lett. {\bf 23} (1972) 142.

\bibitem{HO} H. Hasimoto and H. Ono, J. Phys. Soc. Japan {\bf 33} (1972) 805.

\bibitem{Taniuti} T. Taniuti, Suppl. Proc. Th. Phys. {\bf 55}, 1 (1974).

\bibitem{KT} Y. Kodama and T. Taniuti, J. Phys. Soc. Japan, {\bf 45}, 298 (1978).

\bibitem{CE1} F. Calogero and W. Echkhaus, ``Nonlinear evolution equations, rescalings, model PDEs and their 
integrability: I''Inverse Problems {\bf 3}, 229-262 (1987).
\bibitem{ZS} V. E. Zakharov and A. S. Shabat, Soviet Phys. JETP {\bf 34} (1972) 62.

\bibitem{ZM} V.E. Zakharov and S.V. Manakov, ``Resonance interaction of wave packets'', 
Soviet Physics JETP  42  (1975).

\bibitem{ZK} V. E. Zakharov and E. A. Kuznetsov, Physica D {\bf 18}, 455 (1986).

\bibitem{KdV} D. J. Korteweg and F. de Vries, "On the Change of Form of Long Waves Advancing in a Rectangular Canal, 
and on a New Type of Long Stationary Waves." Philos. Mag. 39, 422-443, 1895.

\bibitem{GGKM} C. S. Gardner, C. S. Greene, M. D. Kruskal, and R. M. Miura, "Method for Solving the Korteweg-de Vries 
Equation." Phys. Rev. Lett. 19, 1095-1097, 1967.
 
\bibitem{Calogero} F. Calogero, ``Why are certain nonlinear PDEs both widely applicable and integrable?'' in 
{\it What is integrability?} 1-62, edited by V.E.Zakharov, Springer, Berlin-Heidelberg (1991).

\bibitem{Hopf-Cole} E. Hopf, Commun. Pure Appl. Math. {\bf 3}, 201 (1950). 
J. D. Cole, Quan. Appl. Math. {\bf 9}, 225 (1951).

\bibitem{ZMNP}
V.E.Zakharov, S.V.Manakov, S.P.Novikov and L.P.Pitaevsky, 
{\it Theory of Solitons. The Inverse Problem Method}, Plenum Press (1984)

\bibitem{CD} F. Calogero and A. Degasperis, ``Spectral Transform and solitons: tools to solve and investigate 
nonlinear evolution equations. Volume one'', North Holland, Amsterdam (1982).

\bibitem{AC}
M.J.Ablowitz and P.C.Clarkson, {\it Solitons, Nonlinear Evolution Equations and Inverse Scattering}, 
Cambridge University Press, Cambridge, 1991.

\bibitem{Konop}
B. Konopelchenko, {\it Solitons in Multidimensions}, World Scientific, Singapore (1993).

\bibitem{CE2} F. Calogero and W. Echkhaus, ``Nonlinear evolution equations, rescalings, model PDEs and their 
integrability: II'', Inverse Problems {\bf 4}, 11-13 (1988).

\bibitem{CDJ1} F. Calogero, A. Degasperis and X.D. Ji, ``Nonlinear Schr\"odinger-type equations from multiscale 
reduction of PDEs. I. Systematic derivation'', J. Math. Phys. {\bf 41}, 6399-6443 (2000). 

\bibitem{CDJ2} F. Calogero, A. Degasperis and X.D. Ji, ``Nonlinear Schr\"odinger-type equations from multiscale 
reduction of PDEs. II. Necessary conditions of integrability for real PDEs'', J. Math. Phys. {\bf 42}, 2635-2652 (2001). 

\bibitem{KM} Y. Kodama and A. V. Mikhailov, ``Obstacles to asymptotic integrability'', in {\it Algebraic aspects of 
integrable systems: in memory of Irene Dorfman}, 173-204, edited by A. S. Fokas and I. M. Gel'fand, Birkh\"auser, 
Boston (1996).

\bibitem{Levi} D. Levi, M. Petrera and C. Scimiterna, ``On the integrability of the discrete nonlinear 
Schr\"odinger equation'', E. P. L. {\bf 84}, 10003 (2008).

\bibitem{Tesi} C. Scimiterna, ``Multiscale techniques for nonlinear difference equations'', PhD 
Thesys, Dept. of Physics, University of Roma3, Roma, Italy (2009).

\bibitem{Scimi} C. Scimiterna, ``Multiscale reduction of discrete Korteweg-de Vries equations'',  
J. Phys. A: Math. Theor. (special issue for the Conference SIDE 8). 

\bibitem{Yam} R. I. Yamilov, ``Symmetries and integrability criteria for differential difference equations''. 
J. Phys. A: Math. Gen., R541-R623 (2006).

\bibitem{SS} V. V. Sokolov and A. B. Shabat, ``Classification of integrable evolution equations'', Sov. Sci. Rev. 
section C4, 221-80 (1984).  

\bibitem{MSY} A. V. Mikhailov, A. B. Shabat and R. I. Yamilov, ``The symmetry approach to the classification of 
nonlinear equations. Complete lists of integrable systems'', Russian Math. Surveys {\bf 42}/4, 1-63 (1987).

\bibitem{Peli} D. Pelinovski, ``Translationally invariant nonlinear Schr\"odinger lattices'', Nonlinearity {\bf 19}, 
2695-2716 (2006).

\bibitem{AL} M. J. Ablowitz and J. F. Ladik, J. Math. Phys. {\bf 17}, 1011 (1979).

\bibitem{Davy} A. S. Davydov, J. Thor. Biol. {\bf 38} 559 (1973).

\bibitem{SSH} W. P. Su, J. R. Schieffer and A. J. Heeger, Phys. Rev. Lett. {\bf 42} 698 (1979).

\bibitem{ELS} J. C. Eilbeck, P. S. Lomdhal, A. C. Scott, Physica D {\bf 16}, 318 (1985). 

\bibitem{HeT} D. Hennig and G. Tsironis, Physics Reports {\bf 307}, 333 (1999).

\bibitem{ABDKS} F. K. Abdullaev, B. B. Baizakov, S. A. Darmanyan, V. V. Konotop and M. Salerno, Phys. Rev. A {\bf 64}, 
043606 (2001).


\bibitem{OJE} M. Oster, M. Johansson and A. Eriksson, ``Enhanced mobility of strongly localized modes in waveguide 
arrays by inversion of stability'', Phys. Rev. E {\bf 67}, 056606 (2003). 

\bibitem{CKKS} C. Claude, Y. S. Kishar, O. Kluth and K. H. Spatschek, ``Moving localized modes in nonlinear 
lattices'', Phys. Rev. B {\bf 47}, 14228-14232 (1993).

\bibitem{Magri} F. Magri, J. Math. Phys. {\bf 19}, 1156 (1978).

\bibitem{GD} I. Gel'fand and I. Dorfman, Funct. Anal. Appl. {\bf 13} (1979); {\bf 14} (1980).

\bibitem{FF} A. S. Fokas and B. Fuchssteiner, Lett. Nuovo Cimento {\bf 28} 299 (1980); Physica {\bf 4D} 
47 (1981). 

\bibitem{SF} P. M. Santini and A. S. Fokas: ``Recursion operators and bi-hamiltonian structures in 
multidimensions.I''; Comm.Math.Phys. 115, 375-419 (1988). 

\bibitem{AKNS} M. J. Ablowitz, D. Kaup, A. C. Newell and H. Segur, Stud. Appl. Math. {\bf 53}, 249 (1974).

\bibitem{AS} U. Aglietti and P. M. Santini, ``Multiscale expansions of difference equations in the small lattice spacing regime.    
Integrability test and numerical confirmations'', preprint (in preparation). 

\bibitem{ASS} U. Aglietti, P. M. Santini and C. Scimiterna, preprint (in preparation).

\end{thebibliography}
\end{document}